\documentclass[aps,twocolumn,prl,amsmath,amssymb,10pt,aps,superscriptaddress]{revtex4-1} 
\usepackage{graphicx}
\usepackage{latexsym}
\usepackage{amsmath}
\usepackage{amsthm}
\usepackage{amssymb}
\usepackage{epstopdf} 
\usepackage{enumerate}
\usepackage{setspace} 
\usepackage{dcolumn}
\usepackage{bm}
\usepackage{setspace} 
\usepackage{color}
\usepackage{youngtab}
\usepackage{subfigure}
\usepackage[colorlinks]{hyperref}
\newcolumntype{L}[1]{>{\raggedright\arraybackslash}p{#1}}
\newcolumntype{C}[1]{>{\centering\arraybackslash}p{#1}}
\newcolumntype{R}[1]{>{\raggedleft\arraybackslash}p{#1}}
\usepackage{wasysym} 
\usepackage{makecell}

\def\vecsign{\mathchar"017E}
\def\dvecsign{\smash{\stackon[-1.95pt]{\vecsign}{\rotatebox{180}{$\vecsign$}}}}
\def\dvec#1{\def\useanchorwidth{T}\stackon[-4.2pt]{#1}{\,\dvecsign}}
\usepackage{stackengine}
\stackMath
\usepackage{tikz}
\usepackage{tikz-3dplot}
\usepackage{braket}
\usepackage{cancel}
\usetikzlibrary{shapes,snakes,arrows,chains,matrix,positioning,scopes,calc}
\usetikzlibrary{decorations.markings}
\tikzset{/pgf/decoration/.cd,
    number of sines/.initial=10,
    angle step/.initial=20,
}
\newdimen\tmpdimen
\pgfdeclaredecoration{complete sines}{initial}
{
    \state{initial}[
        width=+0pt,
        next state=move,
        persistent precomputation={
            \pgfmathparse{\pgfkeysvalueof{/pgf/decoration/angle step}}%
            \let\anglestep=\pgfmathresult%
            \let\currentangle=\pgfmathresult%
            \pgfmathsetlengthmacro{\pointsperanglestep}%
                {(\pgfdecoratedremainingdistance/\pgfkeysvalueof{/pgf/decoration/number of sines})/360*\anglestep}%
        }] {}
    \state{move}[width=+\pointsperanglestep, next state=draw]{
        \pgfpathmoveto{\pgfpointorigin}
    }
    \state{draw}[width=+\pointsperanglestep, switch if less than=1.25*\pointsperanglestep to final, 
        persistent postcomputation={
        \pgfmathparse{mod(\currentangle+\anglestep, 360)}%
        \let\currentangle=\pgfmathresult%
    }]{%
        \pgfmathsin{+\currentangle}%
        \tmpdimen=\pgfdecorationsegmentamplitude%
        \tmpdimen=\pgfmathresult\tmpdimen%
        \divide\tmpdimen by2\relax%
        \pgfpathlineto{\pgfqpoint{0pt}{\tmpdimen}}%
    }
    \state{final}{
        \ifdim\pgfdecoratedremainingdistance>0pt\relax
            \pgfpathlineto{\pgfpointdecoratedpathlast}
        \fi
   }
}

\usetikzlibrary{arrows.meta,bending,decorations.markings,intersections} 
\tikzset{
    arc arrow/.style args={%
    to pos #1 with length #2}{
    decoration={
        markings,
         mark=at position 0 with {\pgfextra{%
         \pgfmathsetmacro{\tmpArrowTime}{#2/(\pgfdecoratedpathlength)}
         \xdef\tmpArrowTime{\tmpArrowTime}}},
        mark=at position {#1-\tmpArrowTime} with {\coordinate(@1);},
        mark=at position {#1-2*\tmpArrowTime/3} with {\coordinate(@2);},
        mark=at position {#1-\tmpArrowTime/3} with {\coordinate(@3);},
        mark=at position {#1} with {\coordinate(@4);
        \draw[-{Stealth[length=#2,bend]}]       
        (@1) .. controls (@2) and (@3) .. (@4);},
        },
     postaction=decorate,
     }
}

\begin{document}

\title{Using disorder to identify Bogoliubov Fermi-surface states}
 \author{ Hanbit Oh}
\affiliation{Department of Physics, Korea Advanced Institute of Science and Technology, Daejeon 305-701, Korea}
\author{ Daniel F.  Agterberg}
 \thanks{agterber@uwm.edu}
\affiliation{Department of Physics, University of Wisconsin, Milwaukee, Wisconsin 53201, USA}
 \author{ Eun-Gook Moon }
 \thanks{egmoon@kaist.ac.kr}
\affiliation{Department of Physics, Korea Advanced Institute of Science and Technology, Daejeon 305-701, Korea}

\begin{abstract}
{We argue that a superconducting state with a Fermi-surface of Bogoliubov quasiparticles, a Bogoliubov Fermi-surface (BG-FS), can be identified by the dependence of physical quantities on disorder. 
In particular, we show that a linear dependence of the residual density of states at weak disorder distinguishes a BG-FS state from other nodal superconducting states. 
We further demonstrate the stability of supercurrent against impurities and a characteristic Drude-like behavior of the optical conductivity. 
Our results can be directly applied to electron irradiation experiments on candidate materials of BG-FSs, including Sr$_2$RuO$_4$, FeSe$_{1-x}$S$_x$, and UBe$_{13}$.}
\end{abstract}
\date{\today}
 
\maketitle

{\it Introduction} : 
Elucidating the role of disorder on interacting quantum many body systems has been a central issue in strongly correlated physics, as manifested in the recent advances in quantum scrambling physics \cite{scramb_0,scramb_1,scramb_2,scramb_4,scramb_5,scramb_6}. 
One important class of interacting systems is strong spin-orbit coupled systems with angular momentum $j=3/2$  \citep{Witczak2,Pr2Ir2O7,Kondo, Nd2Ir2O7, Tokura}. 
A quadratic band touching at the Gamma point in the Brillouin zone naturally hosts a large density of states (DOS), and interaction/disorder effects are significantly enhanced \cite{Moon_NFL,Herbut,QBT_dis1,QBT_dis2,QBT_dis3,QBT_dis4}. 
Not only interesting normal states but also novel superconducting states are predicted, \cite{Agterberg0,Agterberg1,Agterberg2,Venderbos,Savary2,Roy,Nakajima,Kim,Igor,GiBaik,GiBaik3,Witczak4,Timm,Brydon,QBT_QSL}.  
In addition to the traditional gap structures with a full gap, a point-nodal gap, and a line-nodal gap, a Fermi-surface of Bogoliubov (BG) quasiparticles in a superconducting state, a so-called Bogoliubov Fermi-surface (BG-FS) \cite{Agterberg1,Igor_BG2,Hirschfeld_BG1,Hirschfeld_BG2,Kivelson_BG} has been demonstrated.
It has been shown that a BG-FS is topologically protected by a Z$_2$ invariant in centrosymmetric systems with broken time-reversal symmetry\cite{Agterberg3}.
Recently, the role of interactions on such BG-FS has been considered, where it has been shown that such BG-FS states can undergo an instability to a non-centrosymmetric state. \cite{BG_instability1,BG_instability2,BG_instability3}.
It is confirmed that a BG-FS may still survive even with an inversion instability \cite{BG_instability1,Igor_BG1}. 
There have been several candidate materials, including heavy-fermion systems  (URu$_2$Si$_2$, UBe$_{13}$), strontium-based compounds (Sr$_{2}$RuO$_4$, SrPtAs), and doped iron-based superconductors (FeSe$_{1-x}$S$_{x}$), but the existence of a BG-FS has not been demonstrated yet \cite{Schemm,Matsuda,Heffner,Zieve,Sr2RuO4_1,Sr2RuO4_2,HanGyeol,SrPtAs,FeSe1-xSx,Hirschfeld_BG1,Hirschfeld_BG2,FeSe1-xSx_1,FeSe1-xSx_2}.

In the previous literature, several ideas to detect a BG-FS have been
suggested, focusing on the presence of a finite non-zero
DOS at zero energy in the clean limit.
This can be detected through the temperature dependence of single-particle observables such as specific heat or penetration depth. 
However, the properties associated with a non-zero DOS cannot confirm the existence of a BG-FS because line-nodal systems with even infinitesimally low disorder may induce a non-zero DOS \cite{Durst,Graf}. 
Thus, it is highly desired to account for disorder effects on BG-FSs.

In this work, we investigate the role of disorder on a BG-FS and demonstrate that a unique signature allows a BG-FS to be identified from other nodal superconducting states.
In particular, we show that a linear behavior of the residual DOS upon changing disorder and a finite superfluid density are necessary and sufficient conditions of the existence of a BG-FS.
These can be measured by experiments, for example, via electron irradiation experiments. 
We calculate the optical conductivity \cite{optical_Igor,optical_ahn} which is a powerful tool to learn the nature of the superconducting pairing gap even in the presence of disorder. 
Our work reconciles the role of disorder on various superconducting states with different dimensionality of zero-energy excitations and provides a new perspective on realizing exotic superconductivity.

 {\it Model} : 
We consider a model Hamiltonian of a BG-FS.
The total Bogoliubov-de Gennes (BdG) Hamiltonian is given by 
  \begin{eqnarray}
\mathcal{H}_{0}(\vec{k})=\left( \begin{array}{c c}
H_{N}(\vec{k})& \Delta(\vec{k})\\
\Delta^{\dagger}(\vec{k})& -H_{N}^{T}(-\vec{k})
\end{array}\right) 
\end{eqnarray} 
where $\Psi_{\vec{k}}^{T}\equiv (\psi^{T}_{\vec{k}},\psi_{-\vec{k}}^{\dagger})$ is a eight-component Nambu spinor, and $\psi_{\vec{k}}^{T}=(c_{\vec{k},\frac{3}{2}},c_{\vec{k},\frac{1}{2}},c_{\vec{k},-\frac{1}{2}},c_{\vec{k},-\frac{3}{2}})$ is a four-component 
$j = 3/2$ spinor. 
For clarity, we choose a standard Hamiltonian introduced in the previous literature \citep{Agterberg1}.
The kinetic part is described by the so-called, Luttinger Hamiltonian,
\begin{eqnarray}
H_{N}(\vec{k})&=&g_{ij}k^{i}k^{j}-\mu ,
\end{eqnarray}
with 
\begin{eqnarray}
g_{ij}=\frac{\hbar^2}{2m}\Big[\tilde{c}_{0}\delta_{ij}+\sum_{a=1}^{3}  \tilde{c}_{1} \Lambda^{a}_{ij}\gamma_{a}+\sum_{a=4}^{5}  \tilde{c}_{2} \Lambda^{a}_{ij}\gamma_{a}\Big].
\end{eqnarray}
The $3\times 3$ Gell-Mann matrices ($\Lambda^{a}$) and $4\times 4$ Gamma matrices ($\gamma_{a}$) whose explicit forms are introduced in the Supplemental Material (SM).
Three dimensionless parameters $(\tilde{c}_{0},\tilde{c}_{1},\tilde{c}_{2})$ are used with the chemical potential ($\mu$) and the effective mass ($m$).
For pairing, a chiral time-reversal symmetry breaking (TRSB) pairing is chosen, 
\begin{eqnarray}
\Delta (\vec{k})=\Delta_{0}\big[\Gamma_{1}+i\Gamma_{2}\big],
\end{eqnarray}
where the overall pairing amplitude, $\Delta_{0}$, is fixed as a real number and the pairing matrices, $\Gamma_{a}= \gamma_{a}U_T$, are introduced with a $4\times 4$ antisymmetric matrix $U_T=\gamma_{3}\gamma_{1}$. 
For numerical evaluation, we set  $\tilde{c}_{0}=0,\tilde{c}_{1}=\tilde{c}_{2}=m=\mu=1$, where SO(3) symmetry is realized in the normal Fermi-surface. 
Hereafter, our discussion is based on the above  microscopic Hamiltonian unless otherwise stated.

The contours of zero-energy states form a toroid and spheroids in momentum space.  
For $\Delta_{0}\neq 0$, the DOS of clean BG-FSs, $D(E)$, follows the scaling relation, $D(0)\propto |\Delta_{0}|$ and $D(E)-D(0)\propto E^{2}$ in the low energy limit. 
Details on the zero-energy manifold and DOS are explained in SM.

{\it Disorder and Residual density of states} : 
We consider non-magnetic impurities at randomly distributed  positions, $\vec{r}_{a}$. 
Assuming SO(3) rotational symmetry,  the momentum-dependent disorder potential is  
 \begin{eqnarray}
H_{\mathrm{dis}}=\sum_{a=1}^{N_{\mathrm{imp}}}\int_{\vec{k},\vec{k}'}e^{i(\vec{k}'-\vec{k})\cdot \vec{r}_{a}} \Big( \psi_{\vec{k}}^{\dagger}V_{\mathrm{dis}}(\hat{k},\hat{k}') \psi_{\vec{k}'} \Big),
 \end{eqnarray}
with
\begin{eqnarray}
V_{\mathrm{dis}}(\hat{k},\hat{k}')=\sum_{l=0}^{\infty} V_{l}\;P_{l}(\hat{k}\cdot \hat{k}'),\nonumber 
\end{eqnarray} 
 where $N_{\mathrm{imp}}$ is the number of identical impurities, $V_{l}$ is an impurity scattering amplitude.
Hereafter, the short-hand notation, $\int_{\vec{k}}\equiv \int \frac{d^3k}{(2\pi)^3}= \frac{1}{\mathcal{V}}\sum_{\vec{k}}$, is used with a volume of a three-dimensional system, $\mathcal{V}$. 
The Legendre polynomials ($P_l$) capture the angular dependence on the Fermi-surface with an angular momentum quantum number ($l$).
After performing the disorder-average, translation invariance is restored and the Green's function of the disordered BG-FS is modified as, $\mathcal{G}_{\mathrm{dis}}^{-1} (\vec{k},i \omega)= \mathcal{G}_{0}^{-1} (\vec{k},i \omega)- \Sigma_{\mathrm{dis}}(\vec{k},i \omega)$, where $\mathcal{G}_{0}^{-1}(\vec{k},i \omega)=i\omega-\mathcal{H}_{0}(\vec{k})$ is the original Green's function and $\Sigma_{\mathrm{dis}}(\vec{k},i \omega)$ is the disordered self-energy.
In the following, we consider the case with $l=0$ as a proof of concept and consider a dilute limit of disorder, the so-called Born limit. 

Employing the first order Born approximation, the scalar channel contribution to the self-energy is 
\begin{eqnarray}
\Sigma_{\mathrm{dis}}(i\omega)=\frac{r_{0}}{8}\;\int_{\vec{k}}\mathrm{Tr}\Big(\mathcal{G}_{0}(\vec{k},i\omega)\Big),
\end{eqnarray}
with two parameters, $r_{0}\equiv n_{\mathrm{imp}}V_{0}^{2}$, $n_{\mathrm{imp}}\equiv N_{\mathrm{imp}}/\mathcal{V}$.
Note that all channels other than the scalar channel may be neglected and absorbed into the changes of microscopic parameters.
The imaginary part of the self-energy gives the scattering rate, $\Gamma_{\mathrm{dis}}(E + i\eta)=-\mathrm{Im}\Sigma_{\mathrm{dis}}(E+i\eta ) = \frac{r_{0}}{8}\pi D(E)$, via analytic continuation with an infinitesimal convergence parameter, $\eta>0$.
It is evident that there is a non-zero scattering rate, $\Gamma\equiv\Gamma_{\mathrm{dis}}( i\eta)>0$, at zero frequency, as a consequence of the non-zero DOS of a BG-FS.
The scattering rate needs not be solved self-consistently in contrast to superconductors with line nodal gaps where self-consistent calculations are essential. 
The disorder averaged spectral function, $A_{\mathrm{dis}}(\vec{k}, E)=[\mathcal{G}_{\mathrm{dis}}(\vec{k},E+i\eta)-\mathcal{G}_{\mathrm{dis}}(\vec{k},E-i\eta)]/2i$, gives the DOS with disorder scattering potentials,
\begin{eqnarray}
D_{\mathrm{dis}}(E;\Gamma)=-\frac{1}{\pi}\int_{\vec{k}}\mathrm{Tr}\left( A_{\mathrm{dis}}(\vec{k}, E)\right),
\end{eqnarray}
as a function of a scattering rate ($\Gamma$). 
The residual DOS is then defined as a difference between the DOS of dirty and clean systems, $\delta D_{\mathrm{dis}}(\Gamma)= D_{\mathrm{dis}}(0;\Gamma)-D(0)$.

\begin{center}
\begin{figure}[tb]
\vspace{-20pt}
\includegraphics[scale=.46]{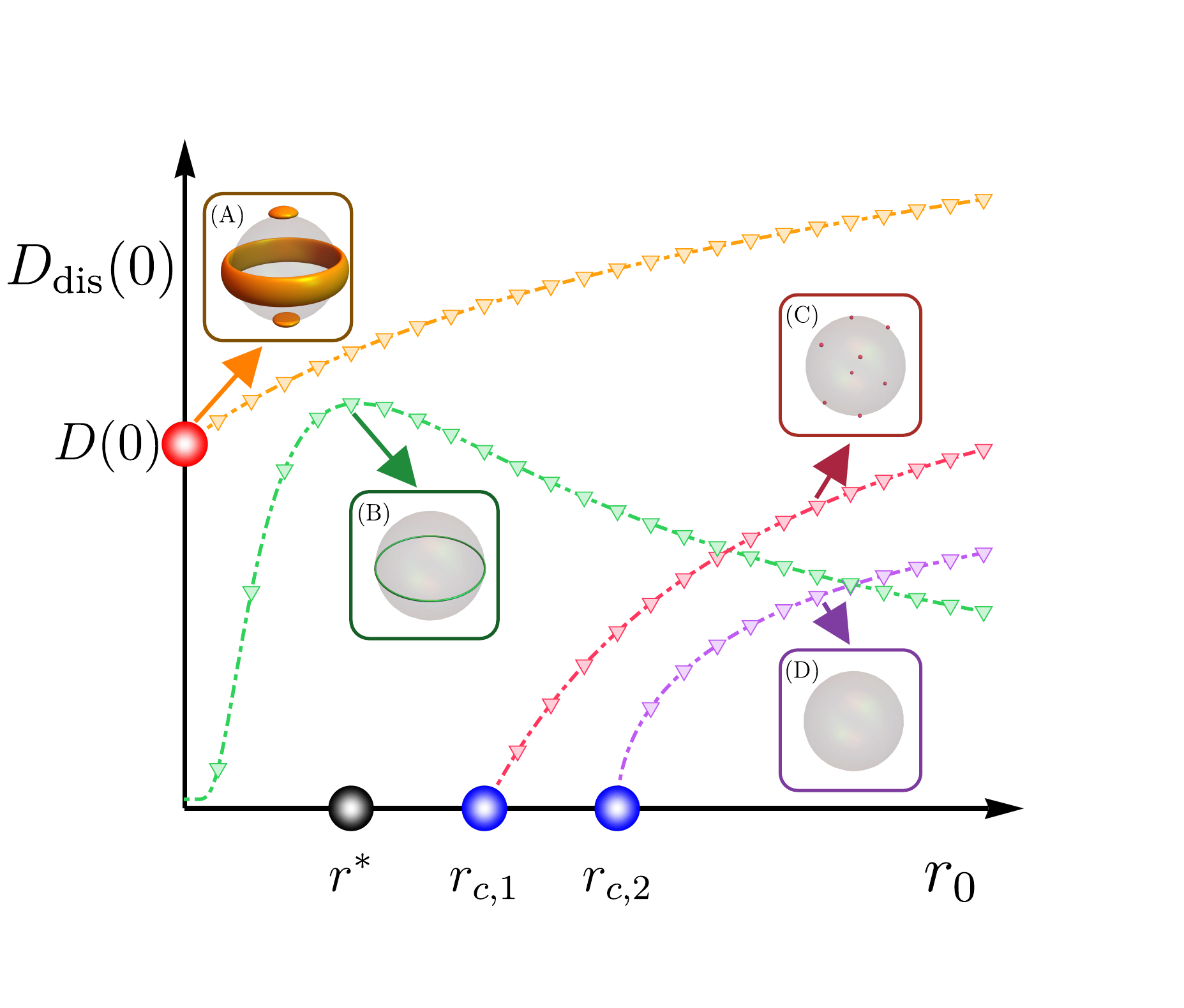}\quad \quad
\vspace{-30pt}
\caption{ Schematic DOS  plot at zero energy ($D_{\mathrm{dis}}(0)$) with various nodal superconductors as a function of impurity strength, $r_{0}=n_{\mathrm{imp}}V_0^2$.
The insets show zero-energy excitations in the momentum space of four different states, BG-FS (A), line-nodal (B), point-nodal (C) and fully gapped (D) superconductors. 
The linear dependence of DOS on $r_{0}$ is a distinctive property of a BG-FS.  
Here, $r^{*}$  is a resonant impurity strength for a line-nodal superconductor \cite{Durst} and $(r_{c,1}, r_{c,2})$ are critical values of $r_{0}$ for point-nodal and fully gapped superconductors, respectively (See SM). 
The functional forms of $\delta D_{\mathrm{dis}}(0;r_{0})=D_{\mathrm{dis}}(0;r_{0})-D(0)$ are tabulated in Table \ref{t1}. 
}\label{f1}
\end{figure}
\end{center}\vspace{-20pt}

In Fig. \ref{f1}, we contrast the residual DOS of a BG-FS with that of superconductors with different nodal structures. 
The $r_{0}$ dependence of the residual DOS is qualitatively different for the different nodal states. 
A few remarks are as follows. 
First, the residual DOS for a BG-FS shows a linear dependence on $r_{0}$. 
To see this, we introduce  a UV energy cut-off ($ \Lambda_{UV}$), for example, the band width, the DOS for a BG-FS is then approximated as 
\begin{eqnarray}
\delta D_{\mathrm{dis}}(\Gamma)=
\int^{\Lambda_{UV}}_{0}\!\!\!\!dED(E)\Big[ \frac{\Gamma/\pi}{E^2+\Gamma^2}-\delta(E)\Big]= a_{0}\Gamma,\ \label{EQapprox}
\end{eqnarray}
at lowest order in $\Gamma/ \Lambda_{UV} \ll 1$. 
For our choice of parameters, we find a linear increase in the residual DOS ($a_{0}>0$) on $\Gamma$. 
Using $\Gamma= \frac{r_{0}}{8}\pi D(0)$,  this implies that the DOS linearly increases upon increasing $r_{0}$. 
We can generalize  the above discussion to a momentum-dependent disorder potential ($l>0$) by including the angle dependent scattering rate, $\Gamma(\vec{k})$, and see$  $ similar results (See SM). 
We remark that the sign of coefficient, $a_{l}$, is not universal but depends on the specific forms of band dispersion and disorder potential.
Second, Eq.(\ref{EQapprox}) may be generalized to systems with different nodal gap structures by considering a generic clean DOS, $D(E)\propto E^{n}$, this allows us to understand the significant differences in Fig. \ref{f1}.
To be specific, for line-nodes, an infinitesimal impurity scattering may induce a zero-energy DOS which follows a non-linear behavior, while it does not affect the DOS unless $r_{0}>r_{c}$ for point-nodes or full-gaps. 
The formulae of DOS as a function of $r_{0}$ are tabulated in Table \ref{t1} and their detailed derivations are explained in SM. 
Thus, we argue that the linear dependence of the residual DOS on impurity scattering is a unique property of BG-FSs.  
Third, the linear dependence of the residual DOS is observable in experiments, for example, in the tunneling conductance between a normal conductor and a BG-FS. 
Standard calculations show that the linear dependence effects are intact even at non-zero temperature, provided that temperature is sufficiently small compared to other energy scales, such as the disorder scattering rate or the
Fermi-energy of Bogoliubov quasi-particles (See SM).

\begin{center}

\begin{table}[t]
\renewcommand{\arraystretch}{1.4}
\begin{tabular}{C{0.15\linewidth}C{0.44\linewidth}C{0.2\linewidth}C{0.15\linewidth}}  \hline \hline
States
&  $ \delta D_{\mathrm{dis}}(r_{0})$&$\mathcal{D}_{s}$&$\mathcal{D}_{D}$ \\ \hline 
(A) &$\propto r_{0}$ &$\Circle$ &$\Circle$ \\
(B)&$\propto \exp [-\frac{r^{*}}{r_{0}}]/r_{0}$ &$\Circle$ &$\times$\\
(C)&$\propto \big[ \frac{1}{r_{c,1}}-\frac{1}{r_{0}}\big]\;\theta(r_{0}-r_{c,1})$ &$\Circle$ &$\times$\\
(D)&$\propto \sqrt{\frac{1}{r_{c,2}}-\frac{1}{r_{0}}}\;\theta(r-r_{c,2})$ &$\Circle$ &$\times$\\
(E) &$\propto r_{0}$ &$\times$ &$\Circle$\\
\hline \hline
\end{tabular}
\caption{
Disorder dependence of physical quantities for different nodal superconducting states: 
(A) BG-FS, (B) line-nodal, (C) point-nodal, and (D) fully gapped superconductors. (E) is for normal metals. 
The functional form, $ \delta D_{\mathrm{dis}}(r_{0})$, superfluid density, $\mathcal{D}_{s}$, and the Drude-weight, $\mathcal{D}_{D}$, in the clean limit are illustrated.}\label{t1}
\end{table}
\end{center}

\begin{center}
\begin{figure}[tb]
\quad \quad \quad (a) \quad \quad \quad \quad \quad  \quad \quad \quad \quad \quad  \quad \ (b)\quad \quad\quad  \\
\includegraphics[scale=.325]{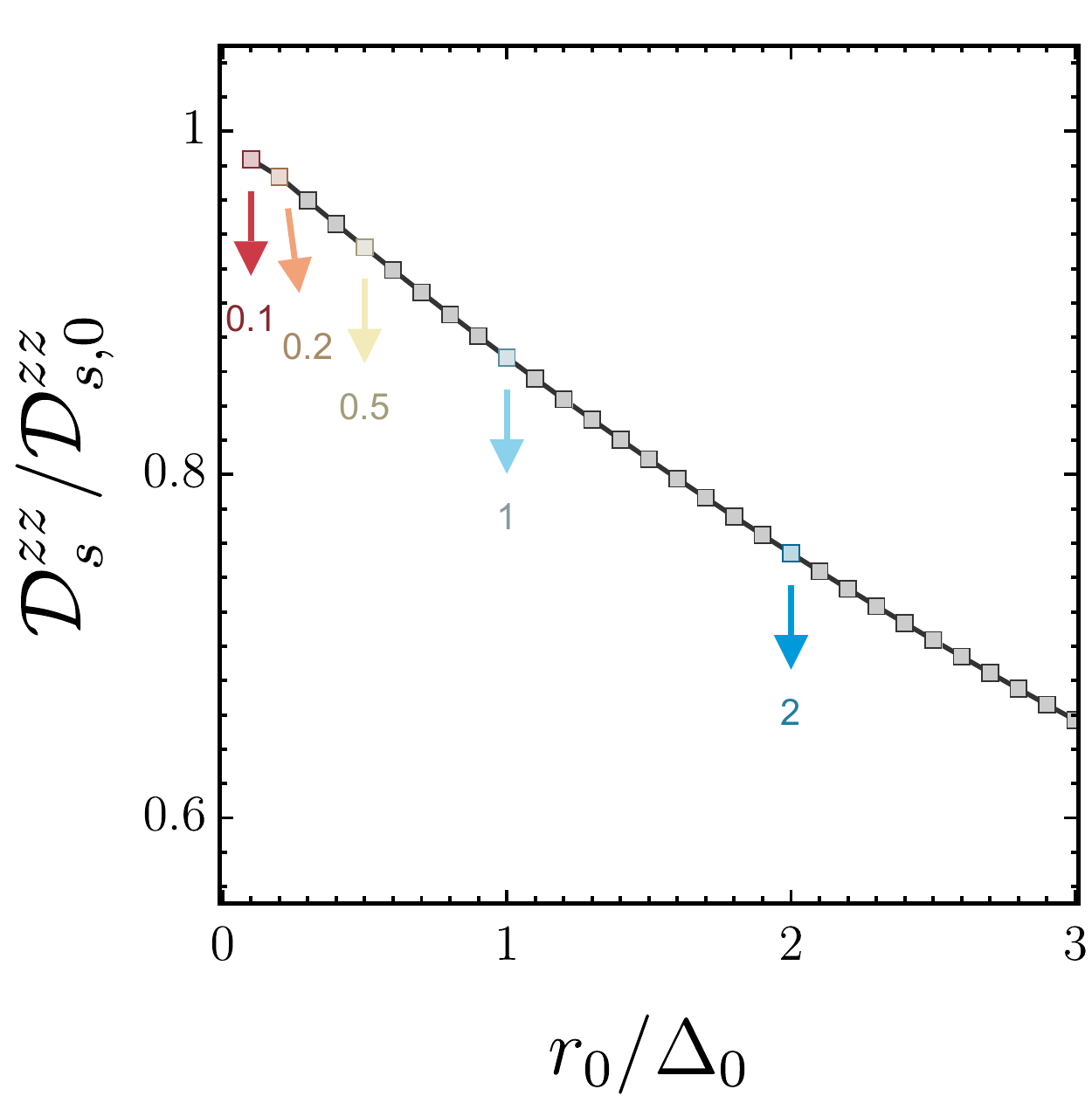} 
\includegraphics[scale=.345]{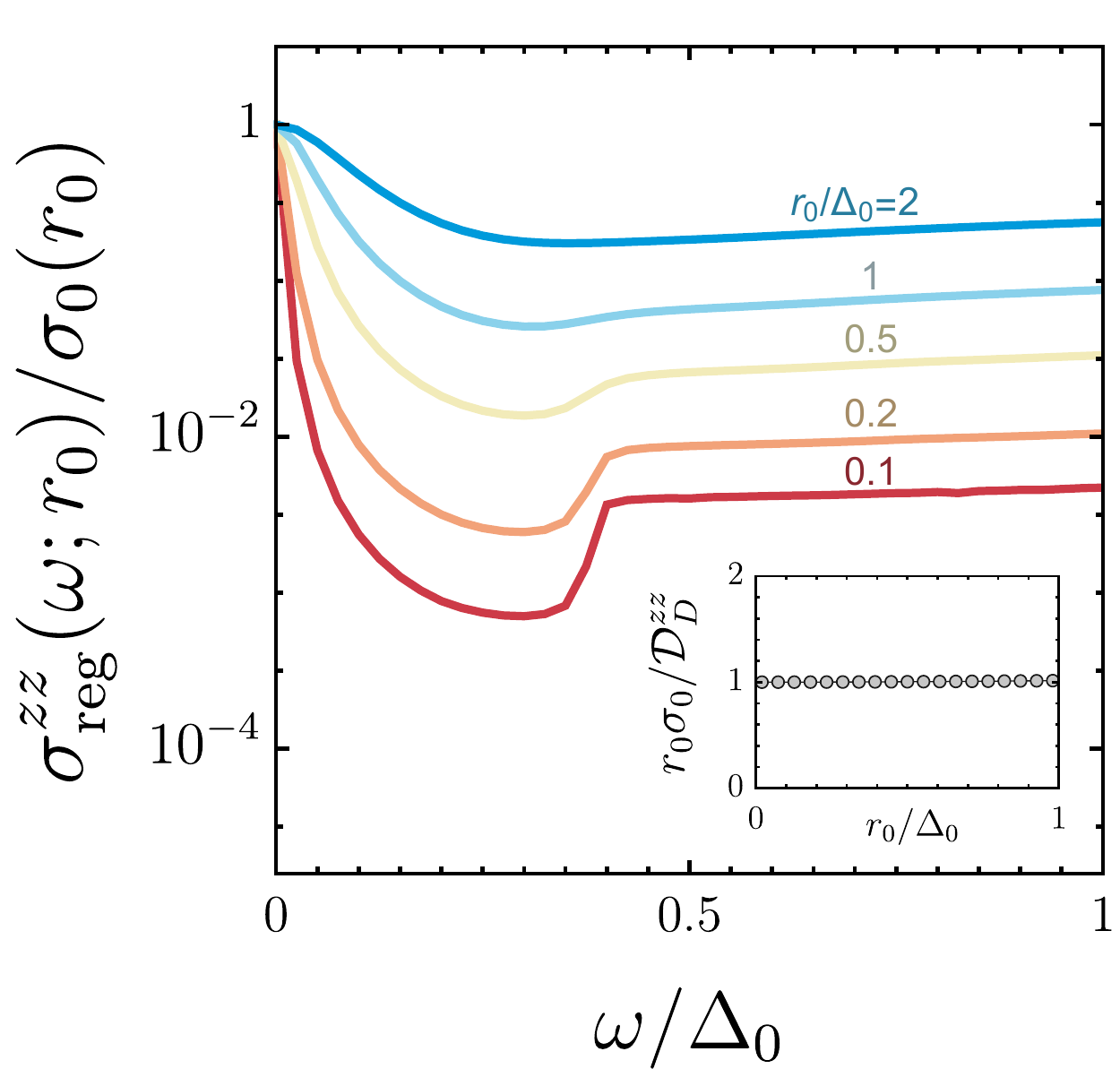}
\caption{(a) The $r_{0}$ dependence of the superfluid weight, $\mathcal{D}_{s}^{zz}(r_{0})$. 
The clean limit of supercurrent is interpolated as a positive value, $\mathcal{D}_{s,0}^{zz}=0.163e^2 \sqrt{m\mu^3}/\hbar^3$. 
(b) The frequency and $r_{0}$ dependence of regular part of conductivity, $\sigma_{\mathrm{reg}}^{zz}(\omega;r_{0})$. 
The different values of $r_{0}/\Delta_{0}=\{0.1,0.2,0.5,1,2\}$ are used and denoted with different colors. 
Inset shows $\sigma_{0}^{zz}(r_{0})=\mathcal{D}_{D}^{zz}/r_{0}$ in the small $r_{0}$ limit with a Drude-weight, $\mathcal{D}_{D}^{zz}=0.168e^2 \sqrt{m\mu^3}/\hbar^3$. 
 } \label{f2} 
\end{figure}
\end{center}\vspace{-20pt}

{\it Optical conductivity }:
Let us consider disorder effects on the optical conductivity of a BG-FS. We focus on two aspects of the optical conductivity: the stability of the supercurrent and the existence of a Drude-like frequency dependence. 
We employ the standard  linear response theory, and the real part of the optical conductivity in the spatially homogeneous limit is, 
\begin{eqnarray}
\mathrm{Re}\sigma^{ij}(\omega)= -\frac{\mathrm{Im}Q^{ij}(\omega+i\eta)}{\omega}+\frac{\mathrm{Re}Q^{ij}(0)}{\pi} \delta(\omega),
\end{eqnarray}
where $Q^{ij}$ is the London response kernel.
The conductivity of superconductors is decomposed into two parts.
The former is called the regular part, $\sigma^{ij}_{\mathrm{reg}}(\omega)=- \mathrm{Im} Q^{ij}(\omega+i\eta)/\omega$,  and the latter is called the singular part from the supercurrent, characterized by the superfluid weight, $\mathcal{D}_{s}^{ij}= \mathrm{Re} Q^{ij}(0)/\pi$.

The current operator is decomposed as the paramagnetic ($p$) and diamagnetic ($d$) parts.
In the Nambu basis ($\Psi_{\vec{k}}$), the zero-momentum current operator reads
\begin{eqnarray}
J^{i}=\int_{\vec{k}}\Psi_{\vec{k}}^{\dagger} \mathcal{J}^{i}(\vec{k})\Psi_{\vec{k}}, \quad  \mathcal{J}^{i}=\mathcal{J}^{i}_{p}+\mathcal{J}^{i}_{d},  
\end{eqnarray}
with 
\begin{eqnarray}
\mathcal{J}_{p}^{i}(\vec{k})=-2\left(\begin{array}{cc}
g_{ij}&0 \\
0 &g_{ij}^{T}
\end{array}\right)k^{j}=-\partial_{i}\mathcal{H}_{0}(\vec{k}) \tau_{z}, 
\end{eqnarray}
and
\begin{eqnarray}
 \mathcal{J}_{d}^{i}(\vec{k})=-2\left(\begin{array}{cc}
g_{ij}&0 \\
0 &-g_{ij}^{T}
\end{array}\right)\!A^{j}= -\partial_{i}\partial_j \mathcal{H}_{0}(\vec{k})A^{j}, 
\end{eqnarray}
where the Hartree unit ($e=\hbar=1$) is used. 
The Pauli-matrix ($\tau_{z}$) acts on the particle-hole space and $\partial_{i}\equiv \partial_{k_{i}}$ is the derivative with respect to the momentum $k^{i}$.
The explicit forms of the paramagnetic and diamagnetic contributions to the London response kernel are 
\begin{eqnarray}
Q^{ij}_{p}(i\omega_{n})&=&T\sum_{ik_n}\int_{\vec{k}}\mathrm{Tr}\left(\mathcal{G}_{\mathrm{dis}}(\vec{k},ik_n) \mathcal{J}_{p}^{i} \mathcal{G}_{\mathrm{dis}}(\vec{k},ik_n+i\omega_n) \mathcal{J}_{p}^{j}\right),\nonumber
\end{eqnarray}
and 
\begin{eqnarray}
Q^{ij}_{d}(i\omega_{n}=0)&=&T \sum_{ik_n}\int_{\vec{k}}\mathrm{Tr}\left(\mathcal{G}_{\mathrm{dis}}(\vec{k},ik_n)\:\partial_{i}\partial_{j}\mathcal{H}_{0}\right),\nonumber
\end{eqnarray} 
respectively.
Note that only the zero-frequency component ($i\omega_n=0$) contributes to the diamagnetic kernel \cite{coleman10}.
In what follows, we focus on the $(i,j)=(z,z)$ component of the conductivity under isotropic disorder ($l=0$) at zero temperature ($T=0$). 

We first consider the singular part of the optical conductivity associated with the supercurrent.  
The superfluid weight is obtained by the relation, $\mathcal{D}_{s}^{zz}(r_{0})= \mathrm{Re}\left[Q^{zz}_{p}(0)+Q^{zz}_{d}(0)\right]/\pi $, whose explicit form is 
\begin{eqnarray}
\mathcal{D}_{s}^{zz}(r_{0})=-\frac{1}{2\pi }\int_{\vec{k}}\int^{\infty}_{-\infty}\! \frac{d\omega}{2\pi }\mathrm{Tr}\left( \big[ \mathcal{G}_{\mathrm{dis}}(\vec{k},i\omega)\mathcal{J}_{p}^{z}(\vec{k}),\tau_{z} \big]^{2} \right). \nonumber
\end{eqnarray}
The commutator in the integrand indicates that $\mathcal{D}_{s}^{zz}(r_{0}) = 0$ if a U(1) symmetric system is considered. 
In Fig. \ref{f2} (a), we illustrate $\mathcal{D}_{s}^{zz}(r_{0})$.  

Our calculations indicate that the supercurrent still survives under weak disorder in a BG-FS.
We note that in contrast to the previous results without disorder, our calculations converge even at $T=0$ due to the scattering rate of the Green's function. 
In the clean limit ($r_{0} \rightarrow 0$), the superfluid density ($\mathcal{D}_{s,0}^{zz}=0.163e^2 \sqrt{m\mu^3}/\hbar^3$) interpolates to a non-zero positive value, which shows the stability of the supercurrent under disorder and temperature. 
These results indicate that the supercurrent survives even with the instability associated with the inversion symmetry breaking in a centrosymmetric BG-FS.   
The superfluid density is naturally suppressed by increasing $r_{0}$, similar to superconducting states with different nodal structures \cite{bcs_2,Maki_1}.   
We stress that the presence of a Fermi-surface of Bogoliubov quasiparticles cannot destroy the supercurrent in contrast to the Landau damping of bosonic excitations in metals.

Next, we calculate the regular part of the optical conductivity of a BG-FS.
After analytic continuation, we find 
\begin{eqnarray}
\sigma_{\mathrm{reg}}^{zz}(\omega; r_{0} )=\!\frac{1}{\omega}\int_{\vec{k}}\!\int^{0}_{-\omega}\!\!\frac{d\nu}{\pi}
\mathrm{Tr}\Big(A_{\mathrm{dis}}(\vec{k},\nu)\mathcal{J}^{z}_{p} A_{\mathrm{dis}}(\vec{k},\nu+\omega)\mathcal{J}^{z}_{p} \Big),    \nonumber
\end{eqnarray}
for $\omega>0$ and $T=0$. 
Here, $A_{\mathrm{dis}}(\vec{k},E)$ is the disorder averaged spectral function of a BG-FS.
In Fig.\ref{f2} (b), $\sigma_{\mathrm{reg}}^{zz}(\omega; r_{0} )$ is plotted as a function of a frequency, $\omega$, and the parameter, $r_{0}$.
A Drude-like behavior near zero frequency with a Lorentzian distribution is obtained, similar to that found in metals. 
In the zero-frequency limit, the DC conductivity, $\sigma_{0}^{zz}(r_{0})\equiv \lim_{\omega\rightarrow 0 }\sigma_{\mathrm{reg}}^{zz}(\omega;r_{0})$, becomes 
\begin{eqnarray}
\sigma_{0}^{zz}(r_{0})=\!\frac{1}{\pi}\int_{\vec{k}}
\mathrm{Tr}\Big(A_{\mathrm{dis}}(\vec{k},0)\mathcal{J}^{z}_{p} A_{\mathrm{dis}}(\vec{k},0)\mathcal{J}^{z}_{p} \Big),
\end{eqnarray}
at zero temperature. 
Note that a non-zero DC-limit conductivity also appears in line-node superconductors, for example, d-wave superconductors \cite{Maki_1,Maki_2,Durst,optical_ahn}, but it is not Drude-like in contrast to a BG-FS where $\sigma_{0}(r_{0})\propto {1/r_{0}}$ is manifested (See Inset of Fig.\ref{f2} (b)). 
In the clean limit, the DC conductivity of a BG-FS diverges and yields a Drude-weight, $\mathcal{D}_{D}^{zz}\neq 0$, 
\begin{eqnarray}
\lim_{r_{0}\rightarrow 0 }\frac{\sigma_{\mathrm{reg}}^{zz}(\omega;r_{0})}{\pi}=\mathcal{D}_{D}^{zz}\;\!\delta(\omega )+\cdots,
\end{eqnarray}
where a non-singular term is omitted in $\cdots$. 
We find $\mathcal{D}_{D}^{zz}=0.168e^2 \sqrt{m\mu^3}/\hbar^3$ for our choice of parameters. 
In the SM,  the DC conductivity of superconducting states with various  nodes and their $\Gamma$ dependencies are shown. 
{Therefore,} The Drude-like behavior is a distinctive feature of a BG-FS.

{\it Discussion and conclusion} : 
Our studies indicate that a BG-FS may be uniquely characterized by the dependence on disorder. 
Thus, we propose electron irradiation experiments can be a powerful tool to identify a BG-FS by observing the linear disorder dependence of the residual DOS and superfluid density, as summarized in Table \ref{t1}.

Our results are directly applicable to experiments. We believe that the candidate materials of BG-FSs such as FeSe$_{1-x}$S$_{x}$ for $x>0.17$  \cite{dopedFeSe} and Th-doped UBe$_{13}$ \cite{Zieve} are promising since a likely intrinsic residual density of states has already been reported to be observed.
We stress that other  experiments such as heat capacity and magneto-optical Kerr effect can be also used to observe the disorder dependence since the residual DOS appears in these observables.

Note that, for simplicity, our calculations mainly focus on the cases with a SO(3) symmetric normal band structure and non-magnetic impurity scattering at zero temperature.
It is straightforward to generalize our calculations to include anisotropic and magnetic impurity scattering and we show in the SM that our main results are not modified.
In particular, the robustness of a BG-FS against disorder is considered, and we prove that the Anderson theorem is violated for a BG-FS in accordance with common wisdom. 
Considering both arbitrary pairings and generic disorder potentials, we quantify the fragility of a superconducting state and generalize the concept of superconducting fitness function and the previous literature \cite{fitness_sigrist,fitness_agterberg,fitness_andersen,fitness_brydon1,fitness_brydon2}. See SM for more information.

The following questions regarding a BG-FS remain to be answered in future research. 
The strong disorder effects on a BG-FS needs to be understood. 
It would be interesting to clarify whether the conventional approach with the non-linear sigma model in symmetry class $\mathrm{D}$ applies \cite{Altland,disorder_Ryu}. 
The verification of the f-sum rule on the linear conductivity of a BG-FS and the generalization to the nonlinear conductivities is also an open question \cite{Gf-sumrule}. 
We believe that our work may raise many interesting future studies and open up new directions to search for exotic superconductivity. \\

{\it Acknowledgement }:
The authors thank Y. Bang and J. Ahn for invaluable discussions and comments.We are particularly grateful to T. Shibauchi for communicating ideas in experiments of doped FeSe. 
H.O. and E.-G.M. are supported by the National Research Foundation of Korea (NRF) grant No. 2019M3E4A1080411, No.2020R1A4A3079707, and No.2021R1A2C4001847. 
D.F.A. was supported by the U.S. Department of Energy, Office of Basic Energy Sciences, Division of Materials Sciences and Engineering under Award DE-SC0021971.

\bibliographystyle{apsrev4-1}
%

\onecolumngrid
\clearpage

\setcounter{equation}{0}
\setcounter{figure}{0}
\setcounter{table}{0}
\setcounter{page}{1}

\maketitle 
\makeatletter
\renewcommand{\theequation}{S\arabic{equation}}
\renewcommand{\thefigure}{S\arabic{figure}}
\renewcommand{\thetable}{S\arabic{table}}

\begin{center}
\textbf{\large Supplemental Material for \\``Using disorder to identify Bogoliubov Fermi-surface states''}
\end{center} 
\begin{center} \vspace{-5pt}
{Hanbit Oh,$^{1}$ Daniel F.  Agterberg,$^{2,\ \textcolor{red}{*}}$ and Eun-Gook Moon$^{1,\ \textcolor{red}{\dagger}}$}\\
\emph{$^{1}$ Department of Physics, Korea Advanced Institute of Science and Technology, Daejeon 305-701, Korea}\\
\emph{$^{2}$ Department of Physics, University of Wisconsin, Milwaukee, Wisconsin 53201, USA}
\end{center}

\section{Fermionic Hamiltonian} \label{ss1}
This section provides detailed information on notations and the Bogoliubov-de Gennes (BdG) Hamiltonian of the $j=3/2$ electronic system. 
\subsection{Notation} \label{ss1_1}
We introduce the five gamma matrices ($\gamma_{a}$), forming a Clifford algebra, $\{\gamma_{a},\gamma_{b}\}=2\delta_{ab}$,
\begin{eqnarray}
\gamma_{1}= \frac{J_{y}J_{z}+J_{z}J_{y}}{\sqrt{3}},\ \gamma_{2}= \frac{J_{z}J_{x}+J_{x}J_{z}}{\sqrt{3}},\  \gamma_{3}= \frac{J_{x}J_{y}+J_{y}J_{x}}{\sqrt{3}},\  \gamma_{4}= \frac{J_{x}^{2}-J_{y}^{2}}{\sqrt{3}},\  \gamma_{5}= \frac{2J_{z}^{2}-J_{x}^{2}-J_{y}^{2}}{3},
\end{eqnarray}
with the angular momentum operators ($J_{i}$),
\begin{eqnarray}
J_{x}=\begin{pmatrix}
0& \frac{\sqrt{3}}{2}&0&0\\
 \frac{\sqrt{3}}{2}&0&1&0\\
 0& 1&0&\frac{\sqrt{3}}{2}\\
0&0&\frac{\sqrt{3}}{2}&0\\
\end{pmatrix},\ \ J_{y}=\begin{pmatrix}
0& -\frac{i\sqrt{3}}{2}&0&0\\
 \frac{i\sqrt{3}}{2}&0&-i&0\\
 0& i&0&-\frac{i\sqrt{3}}{2}\\
0&0&\frac{i\sqrt{3}}{2}&0\\
\end{pmatrix},\ \ J_{z}=\begin{pmatrix}
\frac{3}{2}& 0&0&0\\
0&\frac{1}{2}&0&0\\
 0& 0&-\frac{1}{2}&0\\
0&0&0&-\frac{3}{2}\\
\end{pmatrix}.
\end{eqnarray}
The five quadratic functions are used,
\begin{eqnarray}
d_{1}(\vec{k})=\sqrt{3}k_{y}k_{z}, \ d_{2}(\vec{k})=\sqrt{3}k_{x}k_{z}, \ d_{3}(\vec{k})=\sqrt{3}k_{x}k_{y}, \ d_{4}(\vec{k})=\frac{\sqrt{3}}{2}(k_{x}^2-k_{y}^2), \ d_{5}(\vec{k})=\frac{1}{2}(2k_{z}^{2}-k_{x}^{2}-k_{y}^{2}),
\end{eqnarray}
which may be conveniently written by introducing the Gell-Mann matrices ($\Lambda^{a}$) with
($d_{a}(\vec{k})=\Lambda^{a}_{ij}k^{i}k^{j}$),  
\begin{eqnarray}
\Lambda^{1}=\frac{\sqrt{3}}{2}\begin{pmatrix}
0& 0&0\\
0&0&1\\
0&1& 0
\end{pmatrix},\;
\Lambda^{2}=\frac{\sqrt{3}}{2}\begin{pmatrix}
0& 0&1\\
0&0&0\\
1&0& 0
\end{pmatrix},\;
\Lambda^{3}=\frac{\sqrt{3}}{2}\begin{pmatrix}
0& 1&0\\
1&0&0\\
0&0& 0
\end{pmatrix},\;
\Lambda^{4}=\frac{\sqrt{3}}{2}\begin{pmatrix}
1& 0&0\\
0&-1&0\\
0&0& 0
\end{pmatrix},\;
\Lambda^{5}=\frac{1}{2}\begin{pmatrix}
-1& 0&0\\
0&-1&0\\
0&0& 2
\end{pmatrix}.\;
\end{eqnarray}
We also introduce the unitary part of the time-reversal operator,
\begin{eqnarray}
U_{T}=\gamma_{3}\gamma_{1}=\begin{pmatrix}
0&0&0&1\\
0&0&-1&0\\
0&1&0&0\\
-1&0&0&0\\
\end{pmatrix},\
\end{eqnarray}
to decompose the pairing channels with $\Gamma_a=\gamma_a U_T$.

\subsection{BdG Hamiltonian } \label{ss1_2}
Let us consider a model Hamiltonian which realizes a BG-FS.
The generic form of the BdG Hamiltonian reads
\begin{eqnarray}
H_{\mathrm{BdG}}=\int_{\vec{k}} \Psi_{\vec{k}}\mathcal{H}_{0}(\vec{k})\Psi_{\vec{k}},
\end{eqnarray}
with
  \begin{eqnarray}
\mathcal{H}_{0}(\vec{k})=\left( \begin{array}{c c}
H_{N}(\vec{k})& \Delta(\vec{k})\\
\Delta^{\dagger}(\vec{k})& -H_{N}^{T}(-\vec{k})
\end{array}\right) ,
\end{eqnarray} 
where $\Psi_{\vec{k}}^{T}\equiv (\psi_{\vec{k}}^{T},\psi_{-\vec{k}}^{\dagger})$ is a Nambu spinor,
and $\psi_{\vec{k}}^{T}=(c_{\vec{k},\frac{3}{2}},c_{\vec{k},\frac{1}{2}},c_{\vec{k},-\frac{1}{2}},c_{\vec{k},-\frac{3}{2}})$ is a four-component spinor. 
The low-energy physics of the normal Hamiltonian can be described by the so-called Luttinger Hamiltonian,
\begin{eqnarray}
H_{N}(\vec{k})=g_{ij}k^{i}k^{j}-\mu,\quad  g_{ij}=\frac{\hbar^2}{2m}\left[\tilde{c}_{0}\delta_{ij}+\sum_{a=1}^{3}\tilde{c}_{1}\Lambda^{a}_{ij}\gamma_{a}+ \sum_{a=4}^{5}\tilde{c}_{2}\Lambda^{a}_{ij}\gamma_{a}\right], \label{se8}
\end{eqnarray}
with a chemical potential ($\mu$), an effective mass ($m$), and three dimensionless parameters, ($\tilde{c}_{0},\tilde{c}_{1},\tilde{c}_{2}$).
The parameter $(\tilde{c}_{0})$ quantifies the particle-hole asymmetry, and $|\tilde{c}_{1}-\tilde{c}_2|$ characterizes the cubic anisotropy. 
The doubly degenerate energy eigenvalues of $H_N$ are 
\begin{eqnarray}
E_{0,\pm}(\vec{k})=\left(\frac{\hbar^2 }{2m}\tilde{c}_{0}k^{2}-\mu \right)\pm \frac{\hbar^2}{2m}
\sqrt{\tilde{c}_{1}^{2}\sum_{a=1}^{3}d_{a}(\vec{k})^{2}+\tilde{c}_{2}^{2}\sum_{a=4}^{5}d_{a}(\vec{k})^{2}}.
\end{eqnarray}
For the superconducting state, we choose a time-reversal symmetry breaking (TRSB) chiral pairing,
\begin{eqnarray}
\Delta(\vec{k})=\Delta_{0}(\Gamma_1+i \Gamma_2), \label{se10}
\end{eqnarray}
where the overall pairing amplitude ($\Delta_0$) is a real value.
The zero-energy contours of Bogoliubov (BG) quasiparticles are illustrated in Fig. \textcolor{red}{S1(a)}. 
For numerical evaluation, we take the parameters ($\tilde{c}_{0}=0,\tilde{c}_{1}=\tilde{c}_{2}=m=\mu=1$), unless otherwise stated.

The fermionic Green's function is
\begin{eqnarray}
\mathcal{G}_{0}(\vec{k}, i k_n)  = \left(ik_n - \mathcal{H}_{0}(\vec{k})  \right)^{-1},
\end{eqnarray}
and the spectral weight is obtained via analytic continuation ($ik_n\rightarrow  E\pm i\eta$)
\begin{eqnarray}
A(\vec{k},E) = \frac{\mathcal{G}_{0}(\vec{k},E+i\eta)-\mathcal{G}_{0}(\vec{k},E-i\eta)}{2i},
\end{eqnarray}
with an infinitesimal parameter ($\eta>0$).
The density of states (DOS) of a clean BG-FS is 
\begin{eqnarray}
D(E)=-\frac{1}{\pi}\int_{\vec{k}}\mathrm{Tr}\left( A(\vec{k},E)\right),
\end{eqnarray}
where the short-hand form $\int_{\vec{k}}\equiv \int \frac{d^3k}{(2\pi)^3}\equiv \frac{1}{\mathcal{V}} \sum_{\vec{k}}$ is introduced with a volume of a three-dimensional system, $\mathcal{V}$. 
In Fig. \textcolor{red}{S1(b)}, $D(E)$ is plotted. 
The scaling relation, $D(E)-D(0)\propto E^{2}$ with $D(0)\propto |\Delta_{0}|$, has been found, consistent with the previous literature \cite{s_Timm}. 

\begin{center}
 \begin{figure}[h]
 \centering
\subfigure[]{
\centering
\includegraphics[scale=.35]{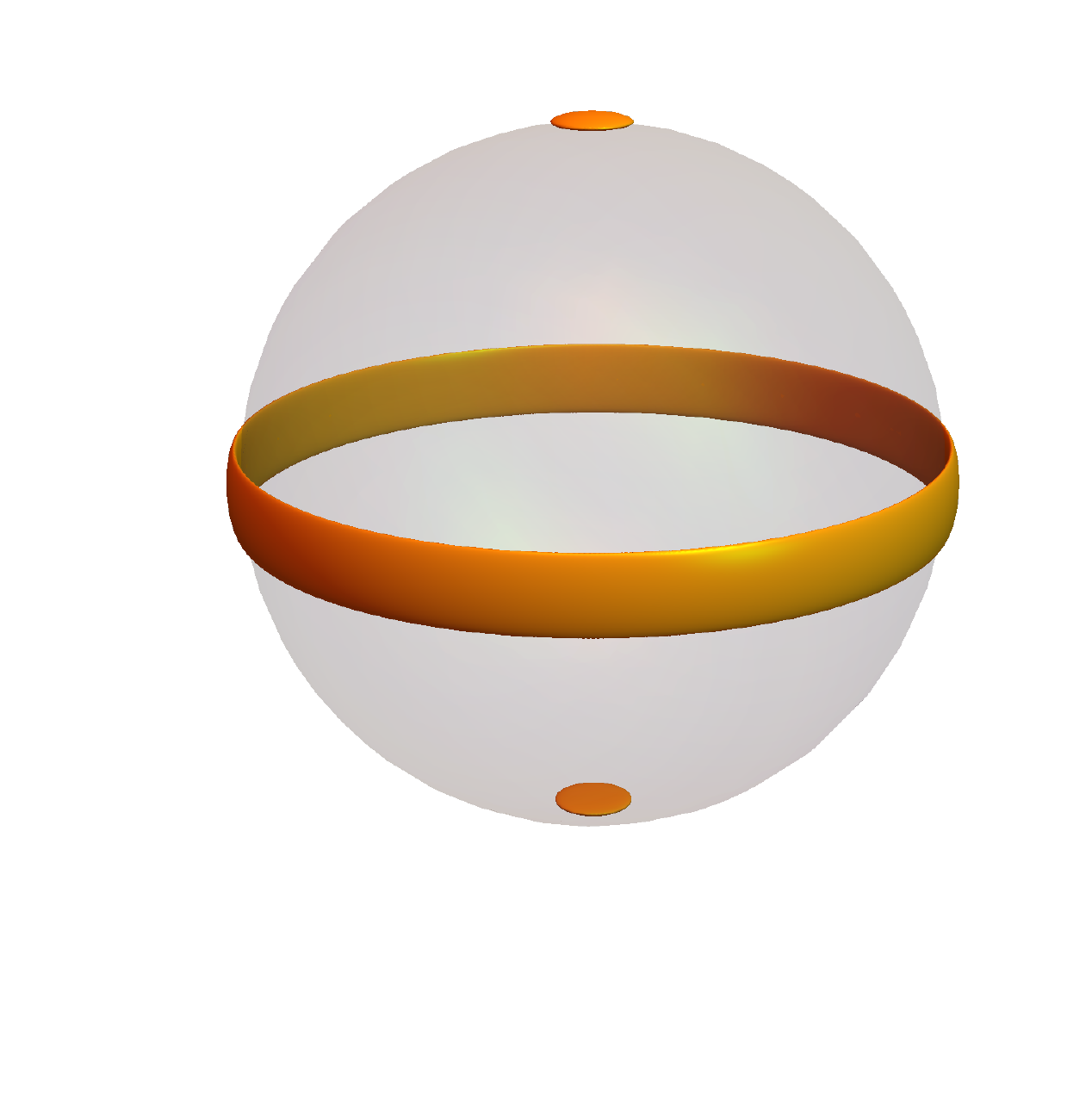}\label{sf1a}
}   \ \ \ 
\subfigure[]{
\includegraphics[scale=.41]{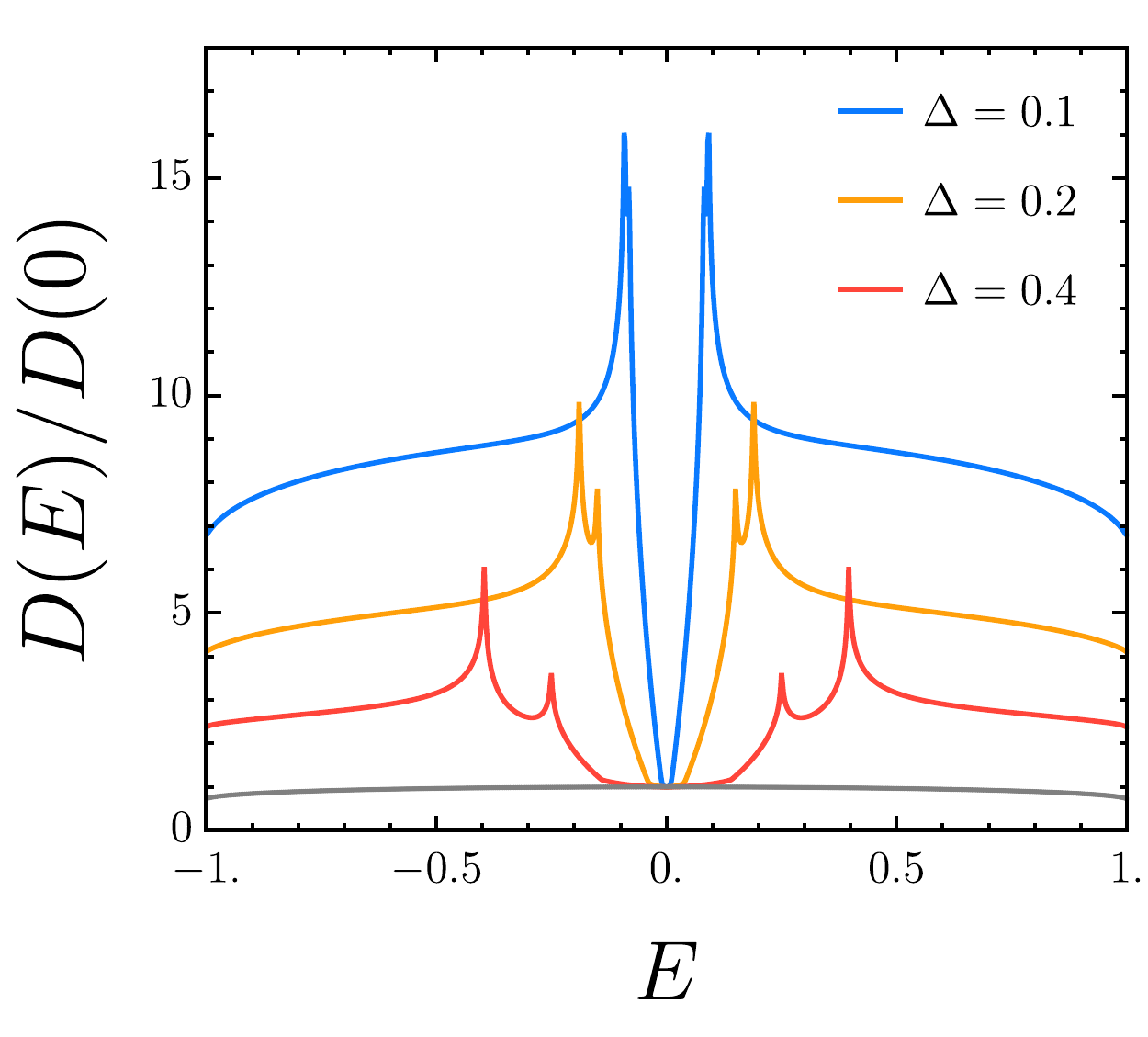}\label{sf1b}
}\vspace{-10pt}
 \caption{(a) Contours of a BG-FS (orange) and a normal Fermi-surface (gray) in the momentum space. 
(b) DOSs of a BG-FS for different pairing amplitudes, $\Delta_{0}=\{0.1,0.2,0.4\}$.
} \label{sf1}
 \end{figure}
\end{center}
\newpage

\section{ Disorder and Residual density of states }\label{ss2}
This section provides the calculation details of the disorder averaged DOS of a BG-FS. 
Let us consider  non-magnetic impurities whose impurity scattering potential is given by 
 \begin{eqnarray}
H_{\mathrm{dis}}=\sum_{a=1}^{N_{\mathrm{imp}}}\int_{\vec{k},\vec{k}'}e^{i(\vec{k}'-\vec{k})\cdot \vec{r}_{a}} \big( \psi_{\vec{k}}^{\dagger}V_{\mathrm{dis}}(\hat{k},\hat{k}') \psi_{\vec{k}'} \big),
\end{eqnarray} 
with 
\begin{eqnarray}
V_{\mathrm{dis}}(\hat{k},\hat{k}')=\sum_{l=0}^{\infty} V_{l}\;P_{l}(\hat{k}\cdot \hat{k}'), 
\end{eqnarray}
where $N_{\mathrm{imp}}$ is the number of impurities, $V_{l}$ is the disorder potential amplitude.
The angular momentum quantum number ($l$) specifies an angular dependence characterized by Legendre polynomials ($P_{l})$. 

In our analysis, we make the following assumptions.
First, we focus on the scalar channel of the self-energy since all other channels may be neglected and absorbed into the changes of microscopic parameters.
Second, we take the aforementioned parameters ($\tilde{c}_{0}=0,\tilde{c}_{1}=\tilde{c}_{2}=m=\mu=1$) for numerical evaluation which realize a SO(3) rotational symmetric normal Fermi-surface.

\subsection{Scattering rate}\label{ss2_1}

The scalar channel of the self-energy under the impurity scattering potential ($l$) is
\begin{eqnarray}
\Sigma_{\mathrm{dis}}(\vec{k},i\omega)=\frac{r_{l}}{8}\;\sum_{\substack{m,m'}}y^{l}_{m}(\hat{k})y^{l*}_{m'}(\hat{k})\Big[\int_{\vec{k}'}y^{l}_{m}(\hat{k}')y^{l*}_{m'}(\hat{k}')\mathrm{Tr}\Big(\mathcal{G}_{0}(\vec{k}',i\omega)\Big)\Big], \label{se14}
\end{eqnarray}
with $r_{l}\equiv n_{\mathrm{imp}}V_{l}^{2}/(2l+1)^2$,  $n_{\mathrm{imp}}=N_{\mathrm{imp}}/\mathcal{V}$ and spherical harmonics, $y^{l}_{m}\equiv \sqrt{4\pi}Y^{l}_{m}$.
The factor $8$ in the denominator manifests the dimension of the BdG Hamiltonian.
The scattering rate is $\Gamma_{\mathrm{dis}}(\vec{k},E+i\eta)=-\mathrm{Im}\Sigma_{\mathrm{dis}}(\vec{k},E+ i\eta)$, and
\begin{eqnarray}
\Gamma_{\mathrm{dis}}(\vec{k},E+i\eta)=\frac{r_{l}}{8}\pi \sum_{\substack{m,m'}}y^{l}_{m}(\hat{k})y^{l*}_{m'}(\hat{k})\Big[\sum_{\alpha}\int_{\vec{k}'}y^{l}_{m}(\hat{k}')y^{l*}_{m'}(\hat{k}')\delta\!\left(E-E_{\alpha}(\vec{k}')\right)\Big],\label{se15}
\end{eqnarray} 
where an analytic continuation ($i\omega \rightarrow E+i\eta$) is performed with introducing band indices, $\alpha$.

For example, the momentum-independent self-energy of isotropic impurities ($\l=0$) is
\begin{eqnarray}
\Sigma_{\mathrm{dis}}(\vec{k},i\omega)=\frac{r_{0}}{8}\;\int_{\vec{k}'}\mathrm{Tr}\Big(\mathcal{G}_{0}(\vec{k}',i\omega)\Big), 
\end{eqnarray}
and its imaginary part is illustrated in Fig. \textcolor{red}{S2}.  The scattering rate of isotropic impurities is
\begin{eqnarray}
\Gamma_{\mathrm{dis}}(E+i\eta)=\frac{r_{0}}{8}\pi D(E),\label{se16}
\end{eqnarray} 
indicating that the impurity scattering is relevant due to the presence of a Fermi-surface of BG quasi-particles, similar to the diffusive mechanism in metals via disorder.

\begin{center}
 \begin{figure}[tb]
\includegraphics[scale=.6]{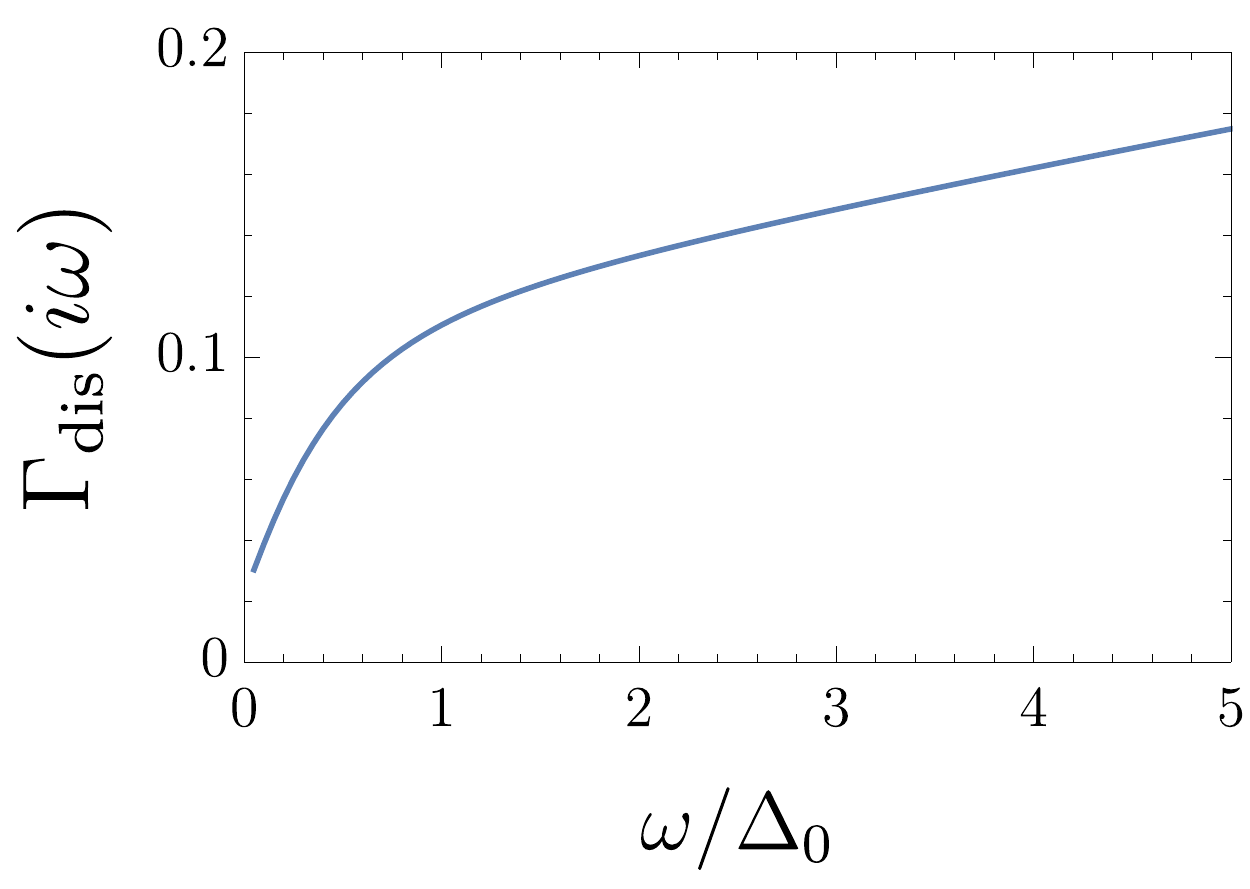}
 \caption{ Functional form $\Gamma_{\mathrm{dis}}(i\omega)\equiv -\mathrm{Im}\Sigma_{\mathrm{dis}}(i\omega)$ with isotropic impurities.
We use the forementioned parameters and $\Delta_{0}=0.2$. 
 In the small $\omega$ limit, the scattering rate is non-zero due to the zero-energy DOS of a BG-FS, $\Gamma_{\mathrm{dis}}(i \eta) = 0.022 r_{0}$.
In the large $\omega$ limit, we find $\Gamma_{\mathrm{dis}}(i\omega)\propto \sqrt{\omega}$, which captures the DOS of quadratic dispersions of the normal state. 
 }\label{sf2}
 \end{figure}
 
\end{center}

\subsection{Disorder averaged DOS }\label{ss2_2}
The disorder averaged Green's function is
\begin{eqnarray}
\mathcal{G}_{\mathrm{dis}}(\vec{k},i k_n)^{-1} = \mathcal{G}_{0}(\vec{k},i k_n)^{-1} - \Sigma_{\mathrm{dis}}(\vec{k},i k_n),
\end{eqnarray} 
and its spectral weight is  
\begin{eqnarray}
A_{\mathrm{dis}}(\vec{k},E)=\frac{\mathcal{G}_{\mathrm{dis}}(\vec{k},E+i\eta)-\mathcal{G}_{\mathrm{dis}}(\vec{k},E-i\eta)}{2i}.
\end{eqnarray}
The disorder averaged DOS at zero-energy is
\begin{eqnarray}
D_{\mathrm{dis}}(0;\Gamma)=-\frac{1}{\pi}\int_{\vec{k}}\mathrm{Tr}\left( A_{\mathrm{dis}}(\vec{k},0)\right)=\frac{1}{\pi}\sum_{\vec{k},\alpha}
\frac{\Gamma(\vec{k})}{E_{\alpha}(\vec{k})^2+\Gamma(\vec{k})^{2}},
\end{eqnarray}
with the zero-energy scattering rate, $\Gamma(\vec{k})\equiv\Gamma_{\mathrm{dis}}(\vec{k},i \eta) $.
In Fig. \textcolor{red}{S3}, the disorder averaged zero-energy DOS, $D_{\mathrm{dis}}(0)$, is plotted as a function of $\Gamma$ and $r_{l}$. 
We stress that the linear dependence of the residual DOS is universal in the small $\Gamma$ limit, and in our model, the residual DOS increases linearly.

It is important to note that the linear dependence of the residual DOS survives at non-zero finite temperature. 
The linear dependence is demonstrated by calculating the tunneling conductance between a normal conductor and superconducting state at zero bias voltage, $G_s=\frac{dI}{dV}\rfloor_{V=0}$, which can be used for the characterization of BG-FSs. 
At zero temperature, this quantity is exactly proportional to the superconducting DOS, $G_{S}\big\rfloor_{V=T=0}\propto D_{\mathrm{dis}}(0;r_{0})$. 
The thermal broadening may cause finite temperature effects and the explicit form is given by, \cite{s_Timm}
\begin{eqnarray}
G_{S}=G_N
\int^{\infty}_{-\infty}\!dE\;\frac{ D_{\mathrm{dis}}(E;r_0)}{D_N}\frac{\partial n_{F}(E/T)}{\partial E},  \label{re1}
\end{eqnarray}
where $n_F(E/T)$ is the Fermi-Dirac distribution function. 
In Fig. \textcolor{red}{S4}, we illustrate the finite temperature results and the linear dependence is clearly shown. 

\begin{center}
 \begin{figure}[b]
 \quad  \quad \quad  (a)\quad\quad \quad \quad \quad \quad \quad \quad \quad  \quad \quad \quad \quad \quad \quad \quad \quad  \quad \quad (b)\\
\includegraphics[scale=.8]{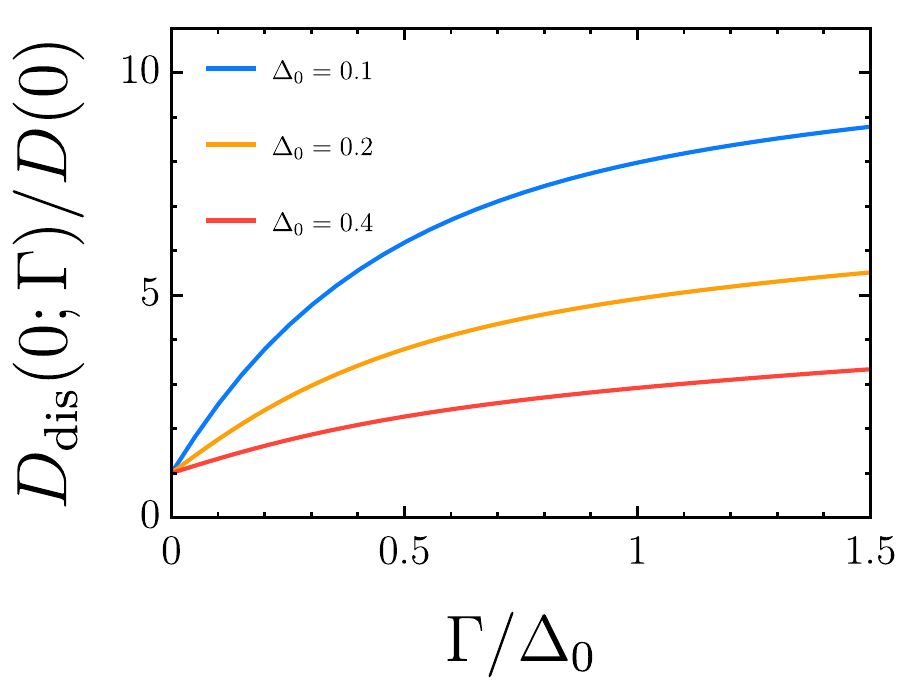}\quad \includegraphics[scale=.55]{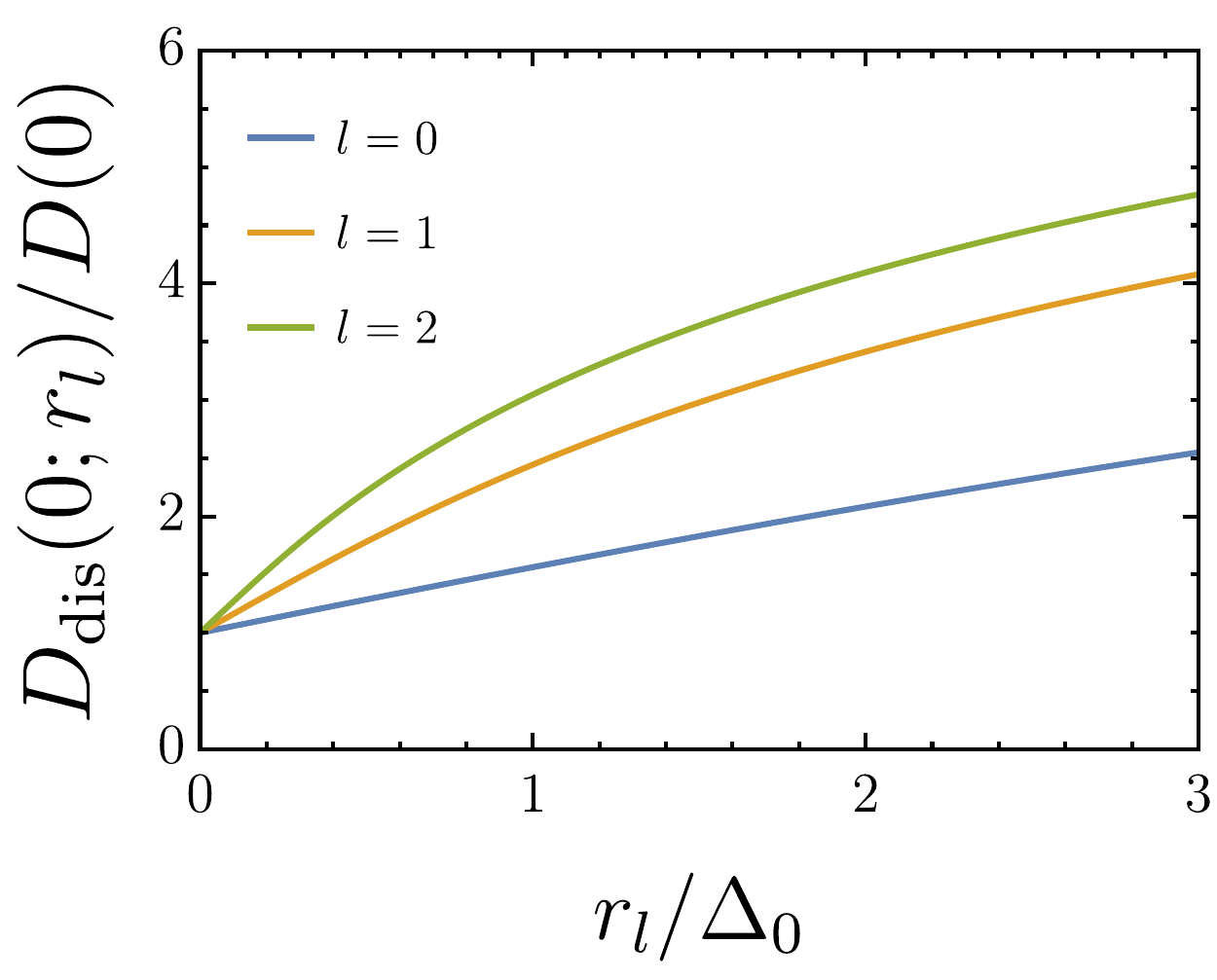}\vspace{-10pt}
 \caption{
(a) $\Gamma$ dependence of the disorder averaged DOS, $D_{\mathrm{dis}}(0;\Gamma)$, under isotropic impurity ($l=0$) with different pairing amplitudes, $\Delta_{0}=\{0.1,0.2,0.4\}$. 
(b) $r_{l}$ dependence of the disorder averaged DOS, $D_{\mathrm{dis}}(0;r_{l})$, with different impurity potentials ($l=0,1,2$). 
Here, we choose a pairing amplitude $\Delta_{0}=0.2$. 
}\label{sf3}
 \end{figure} 
\end{center}
\begin{center}
 \begin{figure}[h]
\includegraphics[scale=.6]{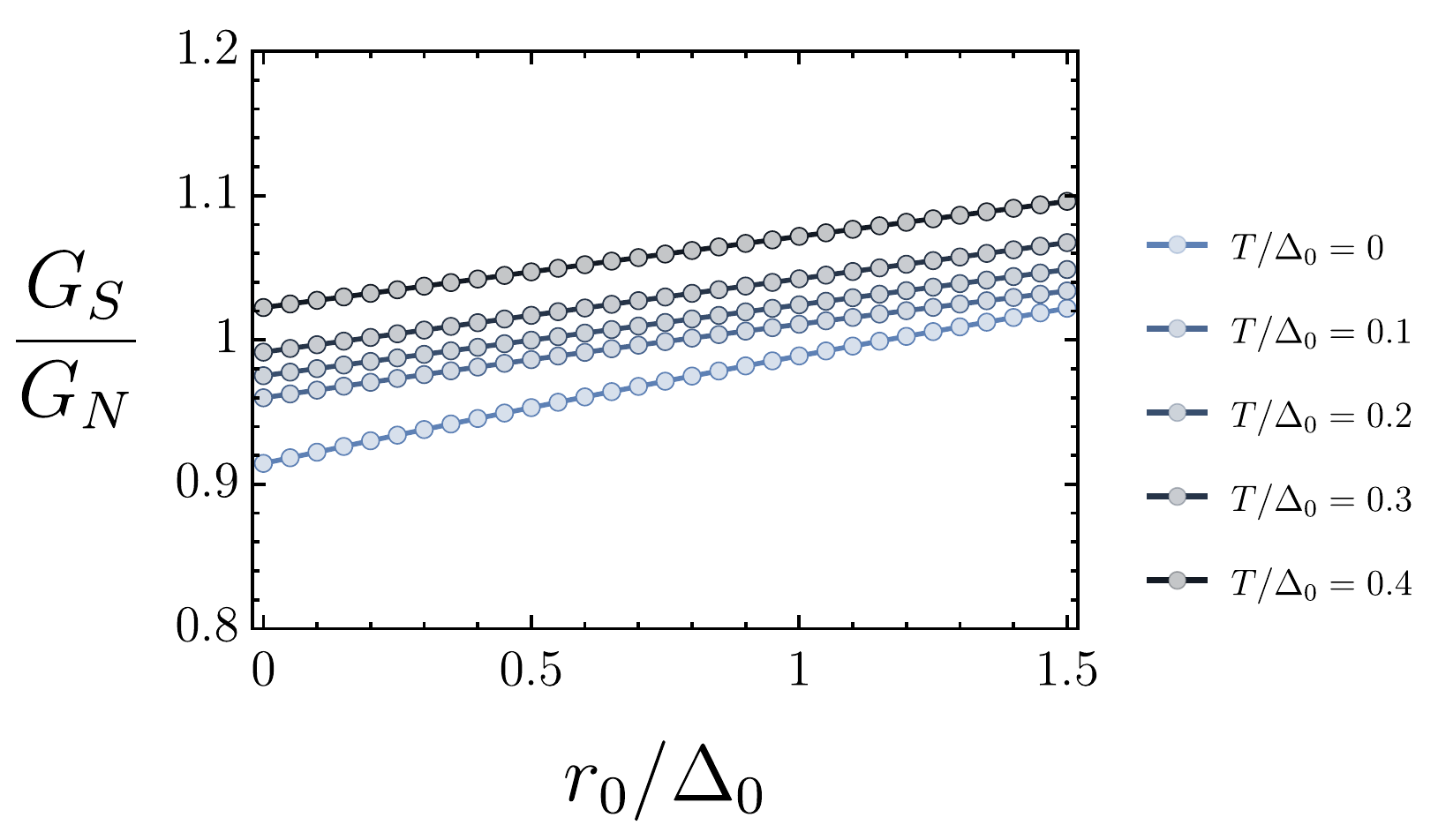}\vspace{-10pt}
 \caption{
The  impurity strength, $r_{0}\equiv n_{\mathrm{imp}}V_0^2$,  dependence of the tunneling conductance ($G_{S}$) between the normal and superconducting states upon changing temperature.
The conductance between the two normal states $(G_{N})$ is introduced.
For all temperatures, $G_S$ is directly proportional to the zero-energy DOS, $D_{\mathrm{dis}}(0;r_0)/D_N$, where $D_N$ is the DOS of the normal state (See Eq.\ref{re1}), and its linear dependence on $r_0$ manifests. 
For a numerical evaluation, we choose a pairing amplitude $\Delta_{0}=1$. 
}\label{sf4}
 \end{figure}
 
\end{center}

\subsection{Comparison with different nodal superconductors} \label{ss2_3}
In this section, we derive the $r_{0}$ dependence of the scattering rate and  zero-energy DOS of superconducting states with different nodal structures. 
Three different types of nodal structures are considered : $i)$ line-node, $ii)$ point-node, $iii)$ full gap.
In these cases, the impurity scattering rate must be calculated self-consistently, while it is determined by clean DOS for a BG-FS, as in Eq. (\ref{se15}).

Adopting the full Born approximation, we introduce the T-matrix, accounting for any number of impurity scattering processes.
The self-energy, $\Sigma_{\mathrm{FBA}}\equiv n_{\mathrm{imp}}T(i\omega_n)$, is defined with the T-matrix ($T$),
\begin{eqnarray}
\quad  T(i\omega_n)=\frac{V_{0}\tau_{z}}{1-V_{0}\tau_{z}\int_{\vec{k}}\mathcal{G}_{\mathrm{dis}}(\vec{k},i\omega_n)},
\end{eqnarray}
where the Pauli-matrix ($\tau_{z}$) acts on the particle-hole space. 
Its scalar channel gives the scattering rate ($\Gamma_{\mathrm{dis}}$),
\begin{eqnarray}
\Gamma_{\mathrm{dis}}(E+i\eta) =-\mathrm{Im}\Sigma_{\mathrm{FBA}}^{0}(E+i \eta), 
\end{eqnarray}
 via analytic continuation. 
In the Born limit ($V_{0}D_{\mathrm{dis}}(0;\Gamma)\ll 1$), including the lowest order diagram is sufficient while all diagrams must be taken into account in the opposite limit, the so-called unitary limit ($V_{0}D_{\mathrm{dis}}(0;\Gamma)\gg 1$).

In the Born limit, the self-energy and the self-consistent equation for the zero-frequency scattering rate are
\begin{eqnarray}
\Sigma_{\mathrm{FBA}}^{0}(E+i \eta)\simeq \frac{n_{\mathrm{imp}}V_{0}^2}{N_{\alpha}}\int_{\vec{k}}\mathrm{Tr}\Big(\mathcal{G}_{\mathrm{dis}}(\vec{k},E+i\eta )\Big) , \label{se21}
\end{eqnarray}
and
\begin{eqnarray}
\Gamma =\frac{ r_{0}}{N_{\alpha}}\pi D_{\mathrm{dis}}(0;\Gamma), \label{se23}
\end{eqnarray}
where $N_{\alpha}$ is a dimension of the BdG Hamiltonian. 
Note that the DOS of a clean system, $D(E)\propto E^{n}$, mainly determines the $\Gamma$ dependence of $D_{\mathrm{dis}}(0;\Gamma)$.
Omitting the band index, $\alpha$, the self-consistent equation becomes
\begin{eqnarray}
1=\frac{r_{0}}{N_{\alpha}}\sum_{\alpha}\int_{\vec{k}}
\frac{1}{E_{\alpha}(\vec{k})^2+\Gamma^{2}}\simeq 
\frac{ r_{0}}{N_{\alpha}}\int^{\Lambda_{UV}}_{0} dE \frac{D(E)}{E^2+\Gamma^{2}},
\end{eqnarray}
with the UV energy cut-off, $\Lambda_{UV}$.
For a BG-FS, Eq. (\ref{se23}) needs not to be solved self-consistently and becomes Eq. (\ref{se15}) at lowest order of $r_{0}$ in the small $r_{0}$ limit. 
In the following,  we consider three different nodes and derive the $r_{0}$ dependence on $D_{\mathrm{dis}}(0)$ shown in Fig.\textcolor{red}{1} and Table.\textcolor{red}{I} in the main text. 

\subsubsection{Line-nodal superconductors}
The DOS of a line-nodal gap ($D(E)\propto |E|$) leads to the self-consistent gap equation  
\begin{eqnarray}
1=\frac{r_{0}}{N_{\alpha}}\int^{\Lambda_{UV}}_{0} dE \frac{D(E)}{E^2+\Gamma^{2}}=\frac{r_{0}}{r^{*}} \log\left( \frac{\Lambda_{UV}}{\Gamma}\right),  
\end{eqnarray}
where $r^{*}$ is a dimensionful parameter.
We find that the scattering rate and zero-energy DOS are
\begin{eqnarray}
 \Gamma=\Lambda_{UV} \exp\left[-\frac{r^{*}}{r_{0}} \right]\propto r_{0}D_{\mathrm{dis}}(0), \quad  D_{\mathrm{dis}}(0)\propto\Lambda_{UV}  \frac{\exp\left[-\frac{r^{*}}{r_{0}} \right]}{r_{0}},
 \end{eqnarray} 
demonstrating that an infinitesimal impurity will induce a residual DOS.

\subsubsection{Point-nodal superconductors}
The DOS of a point-nodal gap ($D(E)\propto E^2$) leads to the self-consistent gap equation,  
\begin{eqnarray}
\frac{1}{r_{0}}\propto \int^{\Lambda_{UV}}_{0} dE \frac{E^{2}}{E^2+\Gamma^{2}}.
\end{eqnarray}
The role of disorder on point-nodes is irrelevant, and the scattering rate is  not affected unless $r_{0}>r_{c}$. 
If the transition to a non-zero DOS is continuous, the critical value ($r_{c,1}$) is determined by taking the $\Gamma\rightarrow 0$ limit,
\begin{eqnarray}
\frac{1}{r_{c,1}}\propto \lim_{\Gamma\rightarrow 0}\int^{\Lambda_{UV}}_{0}  dE  \frac{E^{2}}{E^2+\Gamma^{2}}.
\end{eqnarray}
Comparing the two equations, we find that
\begin{eqnarray}
\frac{1}{r_{c,1}}-\frac{1}{r}\propto \int^{\Lambda_{UV}}_{0}\!\!dE \;\frac{ \Gamma^{2}}{E^{2}+\Gamma^{2}}\simeq \Gamma\left[\frac{\pi}{2}-\frac{\Gamma}{\Lambda_{UV}}\cdots\right], 
\end{eqnarray}
for $\Gamma\ll \Lambda_{\mathrm{UV}}$. 
The scattering rate and zero-energy DOS are
\begin{eqnarray}
\Gamma\propto \Big[\frac{1}{r_{c,1}}-\frac{1}{r_{0}}\Big]\theta \big(r_{0}-r_{c,1}\big),  \quad D_{\mathrm{dis}}(0)\propto \frac{1}{r_{c,1}}\Big[\frac{1}{r_{c,1}}-\frac{1}{r}\Big]\theta\big(r_{0}-r_{c,1}\big),
\end{eqnarray}
near $r_{c,1}\propto \Lambda_{UV}^{-1}$.

\subsubsection{Full gap superconductors }
The DOS of a BCS-type full gap ($D(E)\propto \theta(E-\Delta_{0})E/\sqrt{E^{2}-\Delta_{0}^{2}}$) leads to the self-consistent gap equation,  
\begin{eqnarray}
\frac{1}{r_{0}}\propto \int^{\Lambda_{UV}}_{\Delta_{0}}\!\!dE \;\frac{E}{\sqrt{E^{2}-\Delta_{0}^{2}}}
\frac{1}{E^{2}+\Gamma^{2}}, \quad
\frac{1}{r_{c}}&\propto &\int^{\Lambda_{UV}}_{\Delta_{0}}\!\!dE \;\frac{E}{\sqrt{E^{2}-\Delta_{0}^{2}}}
\frac{1}{E^{2}},
\end{eqnarray}
where $r_{c,2}$ is the critical value of BCS-type superconductor with 
\begin{eqnarray}
\frac{1}{r_{c,2}}-\frac{1}{r_{0}}\propto 
 \frac{\tan^{-1}\left( \frac{\sqrt{\Lambda^2-\Delta^{2}_{0}}}{|\Delta_{0}|}\right)}{|\Delta_{0}|}-
\frac{\tan^{-1}\left( \frac{\sqrt{\Lambda^2-\Delta^{2}_{0}}}{\sqrt{\Delta_{0}^2+\Gamma^2}}\right)}{ \sqrt{\Delta_{0}^2+\Gamma^2}}
 \simeq \left(\frac{\pi}{4|\Delta_{0}|^{3}}\right)\Gamma^{2},
\end{eqnarray}
for $\Gamma\ll \Delta_{0}\ll\Lambda_{UV}$.
The scattering rate and zero-energy DOS are
\begin{eqnarray}
\Gamma\propto \sqrt{\frac{1}{r_{c,2}}-\frac{1}{r_{0}}}\; \theta\big(r_{0}-r_{c,2}\big), \quad D_{\mathrm{dis}}(0)\propto \frac{1}{r_{c,2}}\sqrt{\frac{1}{r_{c,2}}-\frac{1}{r_{0}}}\; \theta\big(r_{0}-r_{c,2}\big),
\end{eqnarray}
near $r_{c,2}\propto |\Delta_{0}|$.

\newpage
\section{Optical conductivity  }\label{ss3}
This section provides the calculation details of the optical conductivity of a BG-FS with impurity scattering.
\subsection{Current operator and London response kernel}    \label{ss3_1}
The gauge field ($\vec{A}$) couples to charged electrons through  minimal coupling ($\vec{k}\rightarrow \vec{k}+\frac{e}{\hbar}\vec{A}$).
Let us consider the normal Hamiltonian ($H_A$) in the presence of an external electromagnetic field,
\begin{eqnarray}
H_{A}=\int d^3 r \big[(\frac{\partial}{i}+\frac{e}{\hbar}A)_{i}\psi_{\vec{r}}\big]^{\dagger}g_{ij}\big[(\frac{\partial}{i}+\frac{e}{\hbar}A)_{j}\psi_{\vec{r}}\big].
\end{eqnarray}
The current operator ($\vec{J}$) is written as 
\begin{eqnarray}
J^{i}= -\frac{\delta H_{A}}{\delta    A^{i}}=i\hbar e\left[\psi_{\vec{r}}^{\dagger}g_{ij}(\partial_{j}\psi_{\vec{r}})-(\partial_{j}\psi^{\dagger}_{\vec{r}})g_{ij}\psi_{\vec{r}}\right]-2e^2\psi_{\vec{r}}^{\dagger}g_{ij}A^{j}\psi_{\vec{r}},
\end{eqnarray}
consisting of the paramagnetic and diamagnetic contributions, $(\vec{J}_p,\vec{J}_d)$. The momentum representations are 
\begin{eqnarray}
J_{p}^{i} (\vec{r})&=&-\frac{ e}{\hbar} \int_{\vec{k},\vec{q}}e^{i\vec{q}\cdot \vec{r}}\psi^{\dagger}_{\vec{k}} \;2g_{ij}\left(k+\frac{q}{2}\right)^{j}\psi_{\vec{k}+\vec{q}}\equiv \int_{\vec{q}}e^{i\vec{q}\cdot\vec{r}}J_{p}^{i}(\vec{q}),
\end{eqnarray}\vspace{-10pt}
\begin{eqnarray}
J_{d}^{i} (\vec{r})&=&-\frac{e^{2}}{\hbar^2}\int_{\vec{k},\vec{q}}e^{i\vec{q}\cdot \vec{r}}\psi^{\dagger}_{\vec{k}}\;2g_{ij}A^{j}\psi_{\vec{k}+\vec{q}}\equiv \int_{\vec{q}}e^{i\vec{q}\cdot\vec{r}}J_{d}^{i}(\vec{q}),
\end{eqnarray}
where the Fourier transformation ($\psi_{\vec{r}}=\int_{\vec{k}}e^{i\vec{k}\cdot\vec{r}}\psi_{\vec{k}}$) is performed with the short-hand notation ($\int_{\vec{k}}\equiv \int \frac{d^3k}{(2\pi)^3}$).  
The zero-momentum parts ($\vec{q}=0$) of the current operators are 
\begin{eqnarray}
J^{i}_{p}(\vec{q}=0)=-\frac{2 e}{\hbar} \int_{\vec{k}} \psi^{\dagger}_{\vec{k}} g_{ij}k^{j}\psi_{\vec{k}}, \quad
\quad J^{i}_{d}(\vec{q}=0)=-\frac{2e^{2}}{\hbar^2}\int_{\vec{k}} \psi^{\dagger}_{\vec{k}} g_{ij}A^{j}\psi_{\vec{k}},
\end{eqnarray}
and generalizing them to the Nambu basis, $\Psi_{\vec{k}}^{T}\equiv (\psi_{\vec{k}},\psi^{*}_{-\vec{k}})^{T}$, gives \cite{s_optical_Igor}
\begin{eqnarray}
J^{i}(\vec{q}=0)=\int_{\vec{k}}\Psi_{\vec{k}}^{\dagger} \mathcal{J}^{i}(\vec{k})\Psi_{\vec{k}}, \quad \mathcal{J}^{i}(\vec{k})=\mathcal{J}^{i}_{p}(\vec{k})+\mathcal{J}^{i}_{d}(\vec{k}),
\end{eqnarray}
with $8\times 8$ matrices, 
\begin{eqnarray}
\mathcal{J}_{p}^{i}(\vec{k})=-\frac{ 2e}{\hbar} \left(\begin{array}{cc}
g_{ij}&0 \\
0 &g_{ij}^{T}
\end{array}\right)\! k^{j}=-\frac{e}{\hbar}\partial_{i}\mathcal{H}_{0}(\vec{k}) \tau_{z}, \quad
 \mathcal{J}_{d}^{i}(\vec{k})=-\frac{2e^{2}}{\hbar^2}\left(\begin{array}{cc}
g_{ij}&0 \\
0 &-g_{ij}^{T}
\end{array}\right)\!A^{j}= -\frac{e^2}{\hbar^2}\partial_{i}\partial_j \mathcal{H}_{0}(\vec{k})A^{j}, 
\end{eqnarray}
where $\tau_{z}$ is the Pauli matrix acting on the particle-hole space.              
The London response kernel ($\dvec{Q}$) is composed of  the paramagnetic and diamagnetic contributions $(\dvec{Q}_p,\dvec{Q}_d)$, whose explicit forms are 
\begin{eqnarray}
Q^{ij}_{p}(i\omega_{n})&=&T\sum_{ik_n}\int_{\vec{k}} \mathrm{Tr}\left(\mathcal{G}_{\mathrm{dis}}(\vec{k},ik_n) \mathcal{J}_{p}^{i}\mathcal{G}_{\mathrm{dis}}(\vec{k},ik_n+i\omega_n) \mathcal{J}_{p}^{j}\right),
\ 
Q^{ij}_{d}=\frac{e^2}{\hbar^2}T\sum_{ik_n}\int_{\vec{k}}\mathrm{Tr}\left(\mathcal{G}_{\mathrm{dis}}(\vec{k},ik_n)\:\partial_{i}\partial_j \mathcal{H}_{0}\right),
\end{eqnarray} 
in the homogeneous limit ($\vec{q}=0$).
Their diagrammatic expressions are illustrated in Fig. \textcolor{red}{S4}. 
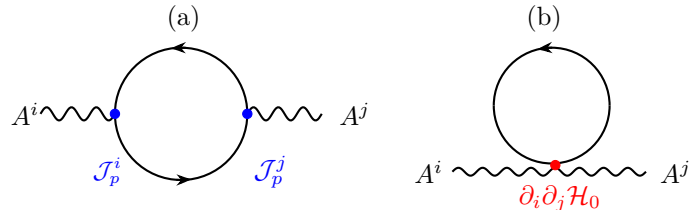
\begin{figure}[b] \vspace{-5pt}
 \ \ \ \  \ \ \ \ \ \ \  (a) \ \ \ \ \ \ \ \ \ \ \ \ \ \ \ \ \ \  \ \ \ \ \ \ \ \ \ \ \ \ \ \ \ \ \ \  (b)    \ \ \ \ \ \ \ \ \ \ \\  \vspace{3pt}
\begin{tikzpicture}[anchor=base,baseline]
\begin{scope}[scale={2.2}] 
\draw[thick,decorate,decoration={complete sines, number of sines=3,amplitude=5pt}](-0.4,0)--(-.85,0);
\draw[thick,decorate,decoration={complete sines, number of sines=3,amplitude=5pt}](0.4,0)--(.85,0);
\draw[thick, arc arrow=to pos 0.55 with length 2mm] (0.4,0) arc (0:180:0.4);
\draw[thick, black,arc arrow=to pos 0.55 with length 2mm] (-0.4,0) arc (180:360:0.4);
\node[left] at(-.8,0.0) {$A^{i}$};
\node[left] at(1.2,0.0) {$A^{j}$};
\node[left,blue] at(-.3,-0.35) {\small $\mathcal{J}_{p}^{i}$};
\node[left,blue] at(.7,-0.35) {$\mathcal{J}_{p}^{j}$};
\filldraw [blue]  (0.4,0) circle [radius=.8pt];
\filldraw [blue]  (-0.4,0) circle [radius=.8pt];
\end{scope}
\end{tikzpicture}\quad
\begin{tikzpicture}[anchor=base,baseline]
\begin{scope}[scale={2.2}]
\draw[thick,decorate,decoration={complete sines, number of sines=8,amplitude=3pt}](-0.6,-0.35)--(0.57,-0.35);
\draw[thick, arc arrow=to pos 0.58 with length 2mm] (0.35,0.05) arc (0:180:0.35);
\draw[thick, black] (-0.35,0.05) arc (180:360:0.35);
\node[left] at(-.6,-0.35) {$A^{i}$};
\node[left] at(.9,-0.35) {$A^{j}$};
\filldraw [red]  (0.022,-0.31) circle [radius=.8pt];
\node[below,red] at(0.05,-0.38) {$\partial_{i}\partial_{j}\mathcal{H}_{0}$};
\end{scope}
\end{tikzpicture} \vspace{-12pt}
\caption{One-loop Feynman diagram of London response kernel ($Q^{ij}_{p},Q^{ij}_{d}$) with $e=\hbar=1$.   
Paramagnetic (a) and diamagnetic (b) contributions are illustrated, respectively. 
Only the paramagnetic London response kernel can carry a non-zero external frequency. }
\label{sf5}

\end{figure}
\subsection{Optical conductivity} \label{ss3_2}
We consider the optical conductivity ($\sigma^{ij}$) within linear response theory,
\begin{eqnarray}J^{i}(i \omega_n)=Q^{ij}(i\omega_n)A^{j}(i\omega_n)=\sigma^{ij}(i\omega_n)E^{j}(i\omega_n). \nonumber
\end{eqnarray} 
The real part of the optical conductivity is obtained via analytic continuation ($i\omega_n\rightarrow \omega+i\eta$), 
\begin{eqnarray}
\mathrm{Re}\;\sigma^{ij}(\omega)=-\frac{\mathrm{Im}\;Q^{ij}(\omega+i\eta )}{\omega}+\frac{\mathrm{Re}\;Q^{ij}(0)}{\pi}\delta(\omega)\equiv \sigma_{\mathrm{reg}}^{ij}(\omega)+\mathcal{D}_{s}^{ij}\;\delta(\omega), 
\end{eqnarray}
 decomposed into two parts. 
The former is called the regular part ($\sigma^{ij}_{\mathrm{reg}}$) and the latter is called the singular part. The singular part determines the superfluid weight ($\mathcal{D}_{s}^{ij}$).
We evaluate the $r_{0}$ dependence of the two parts, focusing on the case with $(i, j) = (z, z)$ at zero temperature.

\subsubsection{ Superfluid weight, $\mathcal{D}^{ij}_s(r_{0})$}
The real part of London response kernel at zero frequency is
\begin{eqnarray}
\mathrm{Re}\big[Q_{p}^{ij}(0)+Q_{d}^{ij}(0)\big]&=&\frac{e^2}{\hbar^2} T\sum_{ik_n} \int_{\vec{k}}\mathrm{Tr}\Big(
\mathcal{G}_{\mathrm{dis}}(\vec{k},ik_n)\left(\partial_{i}\mathcal{H}_{0} \tau_{z}\right)\mathcal{ G}_{\mathrm{dis}}(\vec{k},ik_n)\left(\partial_{j}\mathcal{H}_{0}\tau_z\right)+
\mathcal{G}_{\mathrm{dis}}(\vec{k},ik_n) \partial_{i}\partial_{j}\mathcal{H}_{0}\Big),\\
&=&\frac{e^2}{\hbar^2} T\sum_{ik_n}\int_{\vec{k}} \mathrm{Tr}\Big(
\mathcal{G}_{\mathrm{dis}}(\vec{k},ik_n)\left( \partial_{i}\mathcal{H}_{0}\tau_z\right)\mathcal{ G}_{\mathrm{dis}}(\vec{k},ik_n)\left( \partial_{j}\mathcal{H}_{0}\tau_{z}\right)-\mathcal{G}_{\mathrm{dis}}(\vec{k},ik_n)(\partial_{i}\mathcal{H}_{0})\mathcal{G}_{\mathrm{dis}}(\vec{k},ik_n)(\partial_{j}\mathcal{H}_{0})\Big), \nonumber 
\end{eqnarray}
where the second equality is obtained by using $\mathcal{G}_{\mathrm{dis}}\partial_{i}\partial_{j}\mathcal{H}_{0}=-(\partial_{i}\mathcal{G}_{\mathrm{dis}})\partial_{j}\mathcal{H}_{0}=-(\mathcal{G}_{\mathrm{dis}}\partial_{i}\mathcal{H}_{0}\mathcal{G}_{\mathrm{dis}})\partial_{j}\mathcal{H}_{0}$.
At zero temperature, the Matsubara frequency summation becomes $T\sum_{ik_n}\rightarrow \int^{\infty}_{-\infty}\frac{dk_n}{2\pi}$, thus the superfluid weight is 
\begin{eqnarray}
\mathcal{D}_{s}^{zz}(r_{0})&=&-\frac{1}{2\pi}\int_{\vec{k}}\int^{\infty}_{-\infty}\frac{d k_n}{2\pi} \mathrm{Tr}\Big(\big[ \mathcal{ G}_{\mathrm{dis}}(\vec{k},ik_n) \mathcal{J}^{z}_{p},\tau_z\big]^{2} \Big), 
\end{eqnarray}
where $[\;,\;]$ is a conventional commutator. 
Note that the $r_{0}$ dependence is implicitly in the Green's function, $ \mathcal{ G}_{\mathrm{dis}}(r_{0})$.

\subsubsection{Regular part, $\sigma_{\mathrm{reg}}^{ij}(\omega;r_{0})$} 
The paramagnetic London response kernel is
\begin{eqnarray}
Q^{ij}_{p}(i\omega_{n})&=&\int_{\vec{k}}\int^{\infty}_{-\infty}\frac{d\omega}{\pi}\frac{d\omega'}{\pi} \mathrm{Tr}\left(A_{\mathrm{dis}}(\vec{k},\omega) \mathcal{J}_{p}^{i}A_{\mathrm{dis}}(\vec{k},\omega') \mathcal{J}_{p}^{j}\right)\times \big[\frac{n_{F}(\omega)-n_{F}(\omega')}{\omega-\omega'+i\omega_n}\big],
\end{eqnarray}
where the Lehmann's spectral function, $\mathcal{G}_\mathrm{dis}(\vec{k},ik_n)=\int^{\infty}_{-\infty}\frac{d\omega}{-\pi}\frac{A_\mathrm{dis}(\vec{k},\omega)}{ik_{n}-\omega}$, and the Fermi-Dirac distribution, $n_F(\omega)$, are introduced. 
Taking its imaginary part and performing an analytic continuation ($i\omega_n\rightarrow \nu+i\eta$) leads to
\begin{eqnarray}
\mathrm{Im} Q^{ij}_{p}(\nu+i\eta)&=& \int_{\vec{k}}\int_{-\infty}^{\infty}\frac{d\omega}{(-\pi)}  \mathrm{Tr}\Big(A_\mathrm{dis}(\vec{k},\omega)\mathcal{J}_{p}^{i}A_\mathrm{dis}(\vec{k},\omega+\nu )\mathcal{J}_{p}^{j}\Big)\times \left[n_{F}(\omega)-n_{F}(\omega+\nu)\right], 
\end{eqnarray}
so that the regular part of the conductivity at zero temperature is 
 \begin{eqnarray}
\sigma^{zz}_{\mathrm{reg}}(\nu;r_{0})&=&\frac{1}{\nu}\int_{\vec{k}}\int_{-\nu}^{0}\frac{d\omega}{\pi} \mathrm{Tr}\left( A_\mathrm{dis}(\vec{k},\omega) \mathcal{J}_{p}^{z}  A_\mathrm{dis}(\vec{k},\omega+\nu )\mathcal{J}_{p}^{z}  \right),
 \end{eqnarray}
for $\nu>0$. 
Note that the $r_{0}$  dependence is encoded in the spectral function, $A_\mathrm{dis}(r_{0})$.
At zero-frequency ($\nu\rightarrow 0$), it is simplified as
\begin{eqnarray}
\sigma_{0}^{zz}\equiv \lim_{\nu\rightarrow 0}\sigma_{\mathrm{reg}}^{zz}(\nu;r_{0})=\frac{1}{\pi}\int_{\vec{k}} \mathrm{Tr}\left( A_\mathrm{dis}(\vec{k},0) \mathcal{J}_{p}^{z}  A_\mathrm{dis}(\vec{k},0)\mathcal{J}_{p}^{z}  \right). 
\end{eqnarray}

\subsection{DC conductivity of superconductors with different nodes} \label{ss3_2}
In this section, different types of nodal superconductors are considered to compare with a BG-FS. 
We here focus on the three different pairing states characterized by the following gap matrices,
\begin{eqnarray}
\Delta(\vec{k})&=&\Delta_{\mathrm{B}}\big[d_{1}(\vec{k})+id_{2}(\vec{k})\big] U_T,\quad  \Delta(\vec{k})=\Delta_{\mathrm{C}}\big[d_{4}(\vec{k})+id_{5}(\vec{k})\big]U_T,\quad \Delta(\vec{k})=\Delta_{\mathrm{D}}U_T, \label{se50}
\end{eqnarray}
while using the same parameters for the normal Hamiltonian.
The non-zero pairing amplitudes $(\Delta_{\mathrm{B}},\Delta_{\mathrm{C}},\Delta_{\mathrm{D}})$ are taken to be real numbers and realize a line-node, point-node, and full gap, respectively. 
The zero-energy contours in momentum space are illustrated on the normal Fermi-surface (gray) in Fig. \textcolor{red}{S5 (b-d)}.
While the $\Delta_{\mathrm{B}}$ pairing state has both point- and line- nodes at $\theta_{\vec{k}}=(0,\pi/2)$, the low-energy physics is dominated by the line-node since it contributes much more to the DOS than the point-nodes.
For a $\Delta_{\mathrm{C}}$ pairing, there are eight point-nodes at the $(111)$ direction.

In Fig. \textcolor{red}{S5 (a)}, we plot the DOSs for various states which read
\begin{eqnarray}
D(E;\Delta_{\mathrm{B}})\propto |E|,\quad D(E;\Delta_{\mathrm{C}})\propto E^2,\quad D(E;\Delta_{\mathrm{D}})\propto \frac{E}{\sqrt{E^2-\Delta_{\mathrm{D}}^2}}\;\theta(E-\Delta_{\mathrm{D}}), \label{se51}
\end{eqnarray}
in the low energy limit. 
These scaling behaviors are valid for the common case where the dispersion near the nodes is linear in all three directions.

We investigate the DC conductivity and demonstrate its $\Gamma$ dependence.
Introducing a projection operator, $P_{\alpha}(\vec{k})\equiv|\alpha,\vec{k}\rangle \langle\alpha,\vec{k}|$, of the BdG Hamiltonian with a band indices $(\alpha,\beta)$ the DC conductivity, $\sigma_{0}(\Gamma)$, becomes
\begin{eqnarray}
\sigma^{ii}_{0}(\Gamma)\equiv \lim_{\omega\rightarrow0} \sigma_{\mathrm{reg}}^{ii}(\omega;\Gamma)=\pi \sum_{\alpha,\beta}\int_{\vec{k}}\mathrm{Tr}\big( P_{\alpha}(\vec{k}) \mathcal{J}_{p}^{i}(\vec{k} )  P_{\beta}(\vec{k} )\mathcal{J}_{p}^{i} (\vec{k} ) \big)\times \Big(\frac{\Gamma}{\Gamma^2+E_{\alpha}(\vec{k})^2}\Big)\Big(\frac{\Gamma}{\Gamma^2+E_{\beta}(\vec{k})^2}\Big). 
\end{eqnarray}
Omitting band indices and introducing a UV energy cut-off $(\Lambda_{UV})$, we can employ a conventional dimensional analysis.
For $\Gamma\ll \Delta_{\mathrm{D}} \ll \Lambda_{UV}$, we find that 
\begin{eqnarray}
 \int^{\Lambda_{UV}}_{0}dE E^{n} \Big(\frac{\Gamma}{\Gamma^2+E^{2}}\Big)^{2}\propto \Gamma^{n-1}, \quad 
 \int^{\Lambda_{UV}}_{\Delta_{\mathrm{D}}}dE \frac{E}{\sqrt{E^{2}-\Delta^{2}_{\mathrm{D}}}} \Big(\frac{\Gamma}{\Gamma^2+E^{2}}\Big)^{2}\propto \Gamma^{2}, \quad 
 \end{eqnarray} 
and thus, 
\begin{eqnarray}
\sigma^{ii}_{0}(\Gamma;\Delta_{\mathrm{B}})\propto cons. ,\quad \sigma^{ii}_{0}(\Gamma;\Delta_{\mathrm{C}})\propto \Gamma,\quad \sigma^{ii}_{0}(\Gamma;\Delta_{\mathrm{D}})\propto \Gamma^{2}, \label{se51_2}
\end{eqnarray}
at leading order. 
These scaling relations are numerically confirmed in Fig. \textcolor{red}{S5}.
Therefore we verify that Drude-like behavior, $\sigma^{ii}_{0}(\Gamma;\Delta_{0})\propto \Gamma^{-1}$, manifested in a BG-FS is a unique feature and the non-zero DOS of a BG-FS plays an essential role. 

\begin{center}
 \begin{figure}[h]
  \includegraphics[scale=0.318]{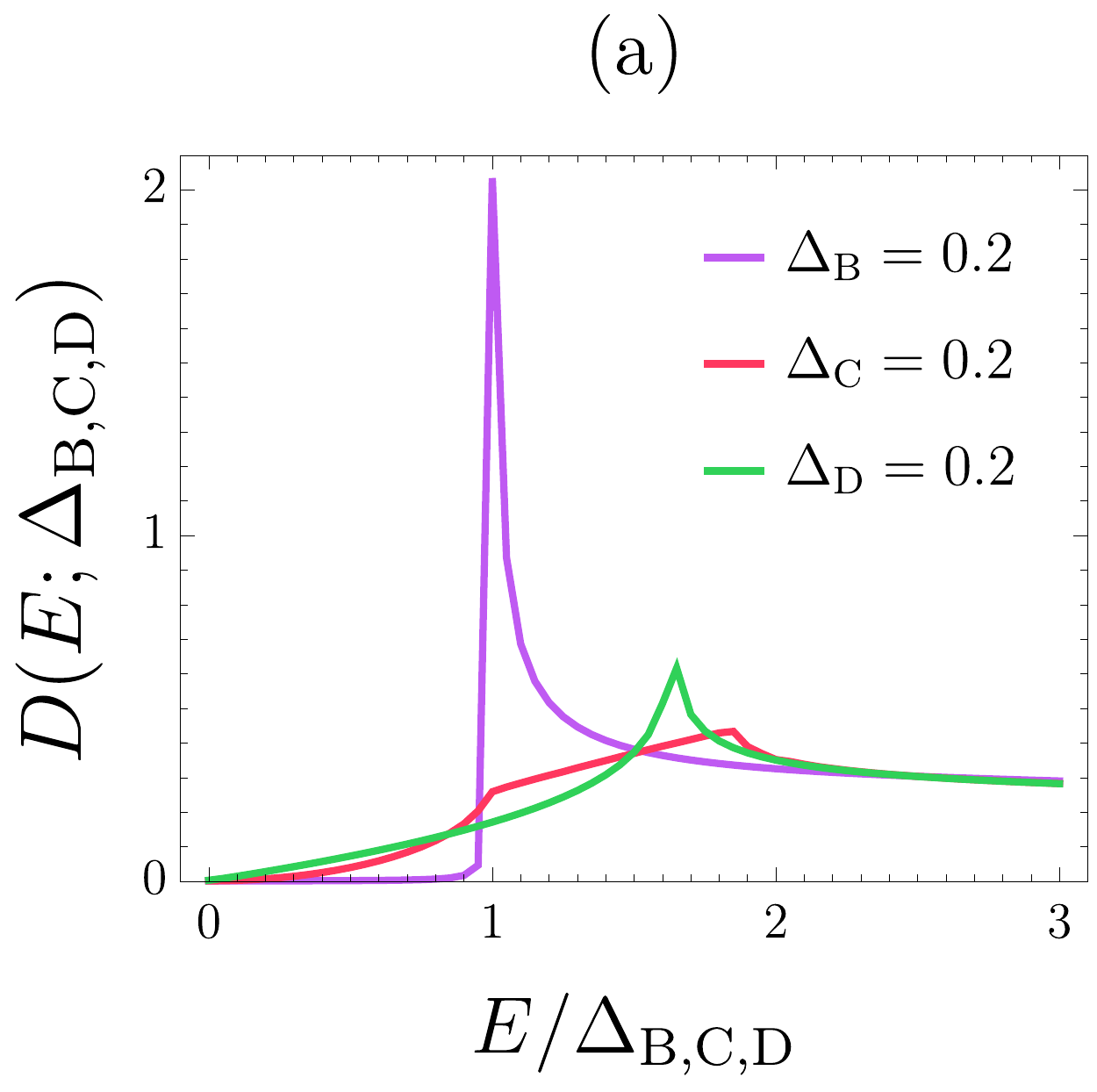}\ \;
  \includegraphics[scale=0.348]{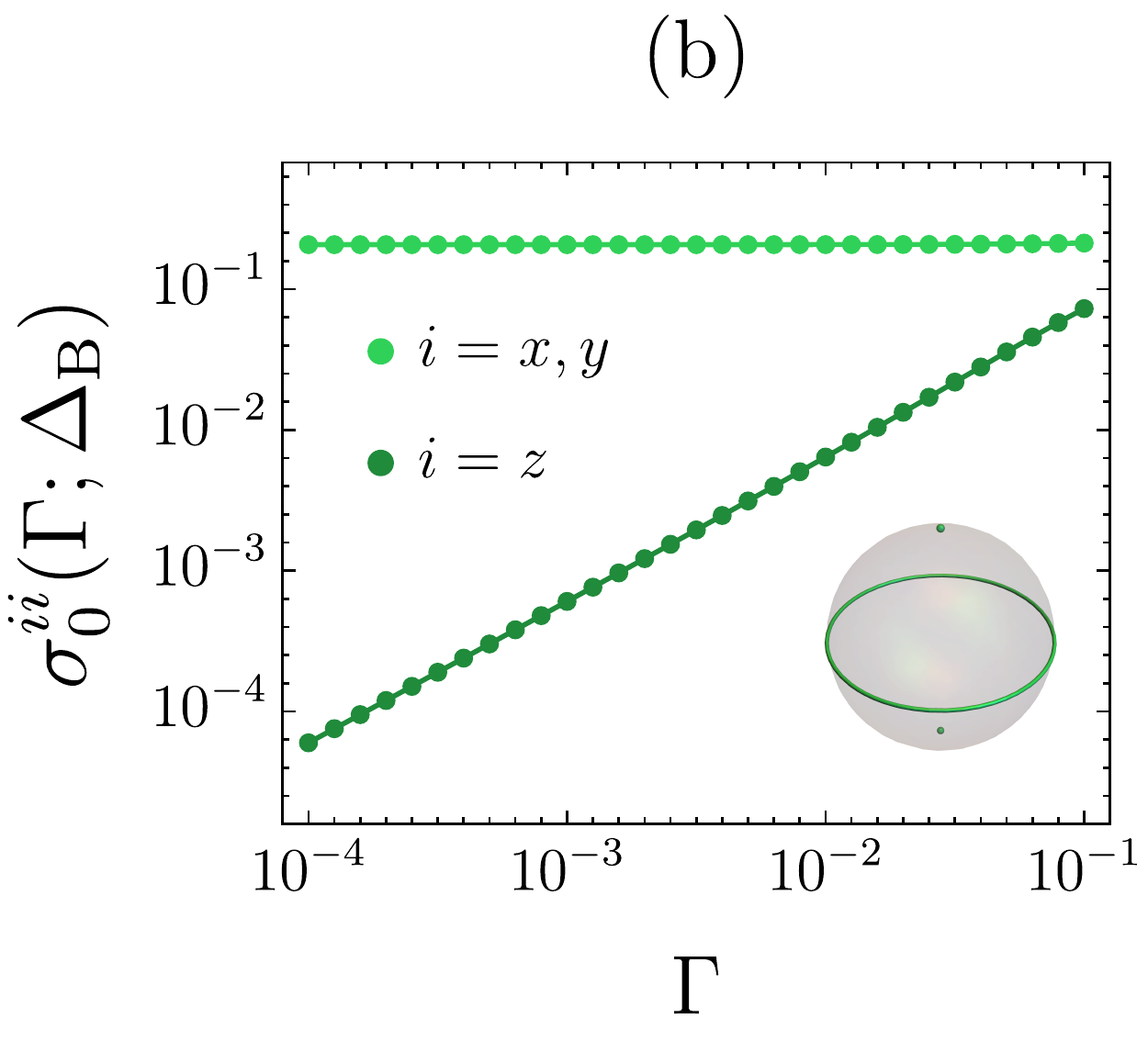} \ 
    \includegraphics[scale=0.348]{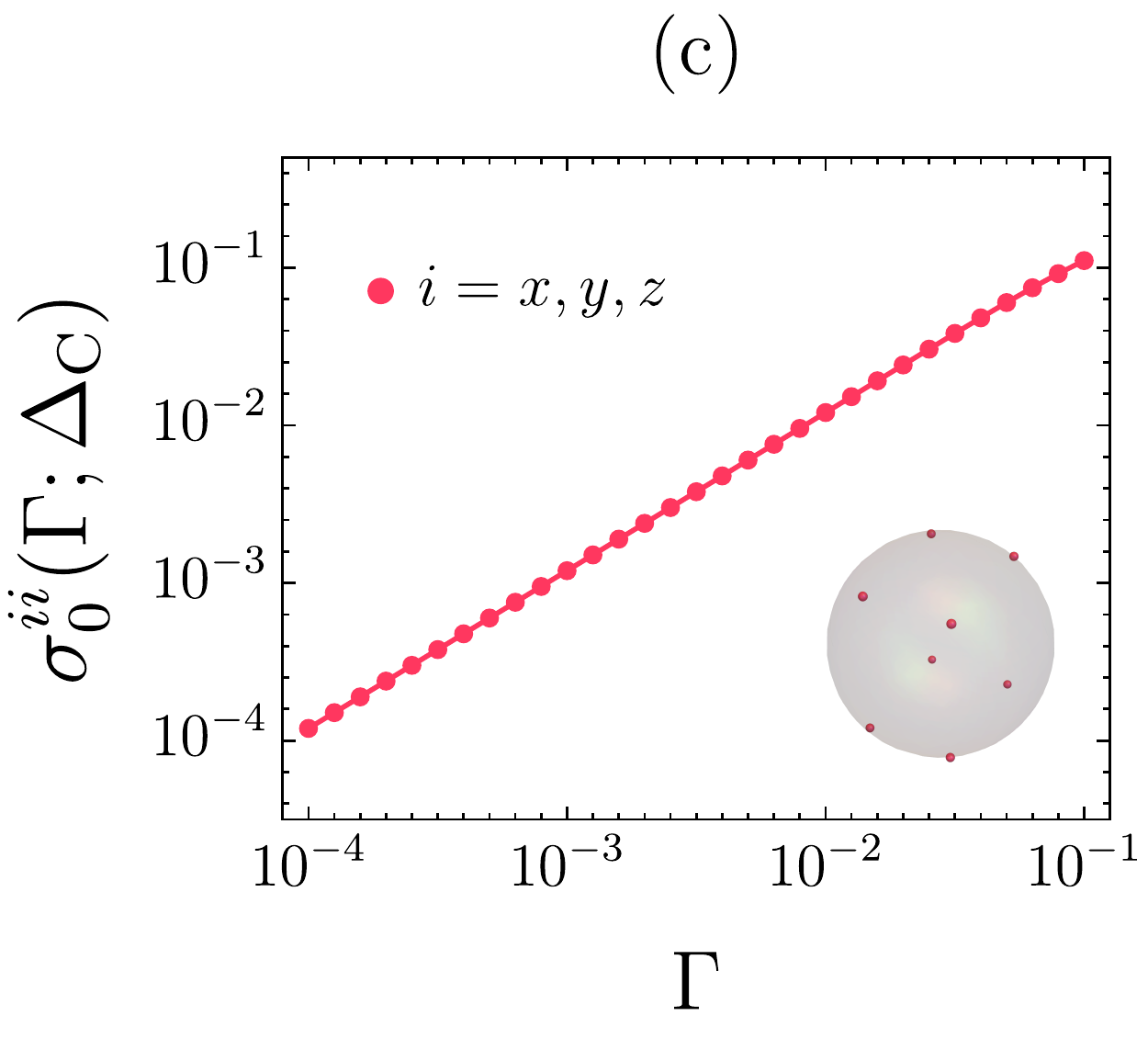}\ 
      \includegraphics[scale=0.348]{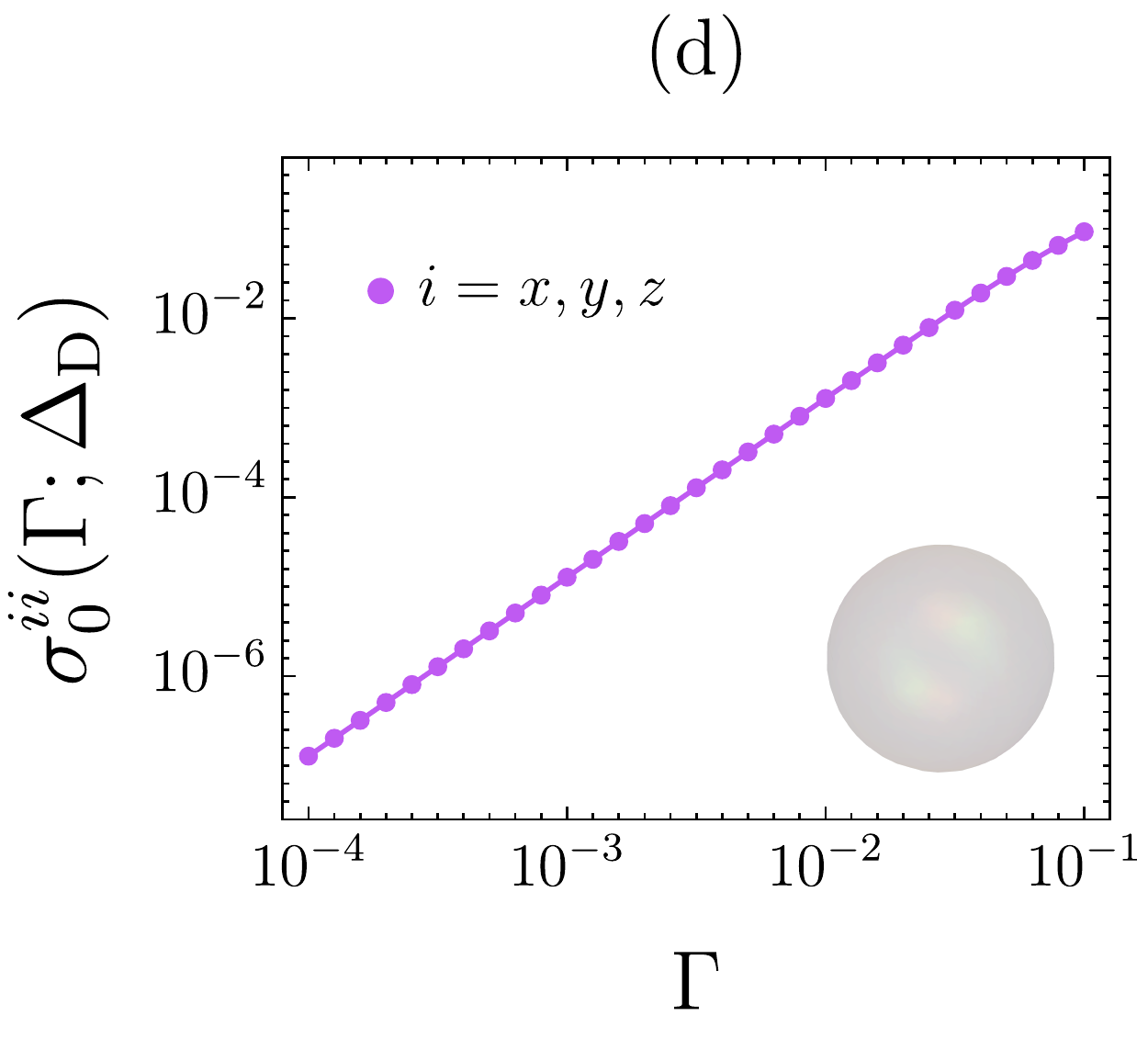} 
\caption{
(a) The DOS of superconductors with three different nodes : line-node ($\Delta_{\mathrm{B}}$), point-node ($\Delta_{\mathrm{C}}$), and full gap ($\Delta_{\mathrm{D}}$), 
where the unit of DOS, $\sqrt{m^3 \mu}/\hbar^3$, is used.
We set $\Delta_{\mathrm{B,C,D}}=0.2$ for each case. 
(b-d) The DC conductivity, $\sigma_{0}^{ii}(\Gamma)$ of different nodal superconductors where the unit of conductivity, $e^2\sqrt{m\mu^{3}}/\hbar^3$, is used. 
They are qualitatively different from that of a BG-FS, $\sigma_{0}^{ii}(\Gamma;\Delta_{0})\propto \Gamma^{-1}$. 
Note that the leading order constant term in $\sigma^{zz}(\Gamma;\Delta_{\mathrm{B}})$ would vanish since the $z$-component of the current operator is zero on the line-node.
Thus the next leading-order contribution has been manifested, $\sigma^{zz}(\Gamma;\Delta_{\mathrm{B}})\propto \Gamma $.
}\label{sf6}
 \end{figure}
\end{center}

\begin{center}

\begin{table}[tb]
\renewcommand{\arraystretch}{1.4}
\begin{tabular}{C{0.08\linewidth}C{0.14\linewidth}C{0.22\linewidth} } \hline \hline
&States&  $\sigma^{ii}_{0}(\Gamma)$\\ \hline 
(A)  & BG-FS&$\propto \Gamma^{-1}$  \\
(B) &Line-node&$\propto \Gamma^{0}$ \\
(C) & Point-node&$\propto \Gamma^{1}$  \\
(D) & Full-gap &$\propto \Gamma^{2}$ \\
\hline \hline
\end{tabular}
\caption{
The leading order scattering rate ($\Gamma$) dependence of DC conductivity of different nodal superconducting states: 
(A) BG-FS, (B) line-nodal, (C) point-nodal, and (D) fully gapped superconductors. 
Only leading-order powers of $\Gamma$ are included. 
 }\label{st1}
\end{table}
\end{center}

\section{  Role of disorder on the critical temperature}\label{ss4}
The suppression of the critical temperature $T_{c}$ in the presence of disorder is one of the key properties of unconventional superconductivity. 
In this section, we evaluate the critical temperature with impurity scattering and propose a parameter to quantify the fragility of a superconducting state, generalizing the concept of a superconducting fitness function \cite{s_fitness_sigrist,s_fitness_agterberg,s_fitness_andersen,s_fitness_brydon1,s_fitness_brydon2}.

\subsection{Two-band Effective Hamiltonian}\label{ss4_1}
Let us introduce a two-band effective Hamiltonian to gain insight into the low energy physics of a BG-FS \cite{s_Agterberg1,s_Agterberg3,s_Venderbos,s_BG_instability1}. 
We consider the case where the weak coupling assumption, $|E_{0,+}-E_{0,-}|_{\vec{k}\in\vec{k}_{F}}\gg\Delta_{0}$, is valid on the normal Fermi-surface \citep{s_Agterberg3},
where $\vec{k}_{F}$ is a Fermi-momentum.
One may construct the two-band effective model by projecting the original BdG Hamiltonian onto either electron or hole bands and ignoring the high energy excitations away from the Fermi-surface. 
To be specific, the effective BdG Hamiltonian projected onto the electron bands reads
  \begin{eqnarray}
\mathcal{H}_{+}(\vec{k})=\left( \begin{array}{c c}
H_{+}(\vec{k})& \Delta_{+}(\vec{k})\\
\Delta^{\dagger}_{+}(\vec{k})& -H_{+}^{T}(-\vec{k})
\end{array}\right) , 
\end{eqnarray} 
where $\tilde{\chi}_{\vec{k}}^{T}=(\chi_{\vec{k}}^{T},\chi^{\dagger}_{-\vec{k}})$ is a Nambu spionor and $\chi_{\vec{k}}^T=(c_{\vec{k},+},c_{\vec{k},-})$ is a two-component pseudospin spinor with indices of pseudospin ($\pm$). 
The explicit forms of normal and pairing parts are
\begin{eqnarray}
H_{+}(\vec{k})&=& h_0(\vec{k})  + \vec{h}(\vec{k})\cdot\vec{ \sigma}, \quad \Delta_{+}(\vec{k})= \Delta_0 \psi_{+}(\vec{k})i \sigma_{y},
\end{eqnarray}
with
\begin{eqnarray}
h_0(\vec{k})&=& E_{0,+}(\vec{k})-\frac{\Delta_{0}^{2}}{2|d(\vec{k})|^3} \left(2d(\vec{k})^2-d_{2}(\vec{k})^{2}-d_{3}(\vec{k})^{2}\right),\quad
\vec{h}(\vec{k})=\frac{\Delta_{0}^{2}}{|d(\vec{k})|}\left(\frac{d_1 (\vec{k})}{\sqrt{3}},\frac{d_2 (\vec{k})}{\sqrt{3}},\frac{|d(\vec{k})|-4d_5 (\vec{k})}{3}\right), \label{se45}
\end{eqnarray}
where $\vec{h}(\vec{k})$ is a pseudomagnetic field and $|d(\vec{k})|$ is a magnitude of a five-dimensional vector, $d_{a}(\vec{k})$. 
The pairing gap function of a pseudospin singlet is $\psi_{+}(\vec{k}) =(d_{1}(\vec{k})+id_{2}(\vec{k})) /|d(\vec{k})|$ indicating a TRSB chiral d-wave.
The Green's function of the effective BdG Hamiltonian is $\mathcal{G}_{+}(\vec{k},ik_n)=(ik_n-\mathcal{H}_{+}(\vec{k}))^{-1}
$. 

In this section, we consider the not only momentum- but also spin dependence of disorder by generalizing the form of the disorder potential.
For example, under SO(3) symmetry, the disorder potential is
\begin{eqnarray}
 \tilde{H}_{\mathrm{dis}}&=&\sum_{a=1}^{N_{\mathrm{imp}}}\int_{\vec{k},\vec{k}'}e^{i(\vec{k}'-\vec{k})\cdot \vec{r}_{a}}\big(\chi_{\vec{k}}^{\dagger} V_{\mathrm{dis}}(\hat{k},\hat{k}')\chi_{\vec{k}'}\big), 
 \end{eqnarray} with
\begin{eqnarray}
V_{\mathrm{dis}}(\hat{k},\hat{k}')&=&\sum_{l}P_{l}(\vec{k},\vec{k}')\Big[ v_{l}+u_{l}\;\vec{S}\cdot \vec{\sigma}\Big], 
\end{eqnarray}
where the quantum number ($l$) specifies the momentum dependence with the corresponding Legendre polynomial ($P_{l}$).
The coupling constants of spin-independent and dependent channels ($v_{l}$, $u_{l}$) and spin of magnetic impurities ($\vec{S}$) are introduced. 
In the Nambu basis ($\tilde{\chi}_{\vec{k}}$), the disorder potential is 
\begin{eqnarray}
\tilde{V}_{\mathrm{dis}}(\hat{k},\hat{k}')=\left(\frac{\tau_{0}+\tau_{z}}{2} \right)\otimes V_{\mathrm{dis}}(\hat{k},\hat{k}')-\left(\frac{\tau_{0}-\tau_{z}}{2} \right)\otimes V_{\mathrm{dis}}^{*}(\hat{k},\hat{k}'),
\end{eqnarray}
where $\tau_{a}$ is the Pauli matrix acting on particle-hole space.

\subsection{Renormalized parameters with disorder }\label{ss4_2}
Accounting for the impurity scattering process, the self-energy  of a singlet pairing state ($\psi_{s}(\vec{k})$) is
\begin{eqnarray}
\Sigma(\vec{k},ik_{n})=n_{\mathrm{imp}}\int_{\vec{k}'}\;\tilde{V}_{\mathrm{dis}} (\hat{k},\hat{k}')
\mathcal{G}_{+}(\vec{k}',ik_{n})\tilde{V}_{\mathrm{dis}}(\hat{k},\hat{k}'),
\end{eqnarray}
which can be written into its frequency and pairing channels, denoted by $(\omega ,\Delta)$, as
\begin{eqnarray}
\Sigma_{\omega}(\vec{k},ik_{n})=\lim_{\Delta_{0}\rightarrow 0}\int_{\vec{k}'}\frac{n_{\mathrm{imp}}}{2}\;\mathrm{Tr}\Big(V_{\mathrm{dis}} (\hat{k},\hat{k}')
G(\vec{k}',ik_{n})V_{\mathrm{dis}}(\hat{k},\hat{k}')\Big),
\end{eqnarray}
and
\begin{eqnarray}
 \frac{\Sigma_{\Delta}(\vec{k},ik_{n})}{\Delta_{0}}=-\lim_{\Delta_{0}\rightarrow 0}\;\frac{n_{\mathrm{imp}}}{2\Delta_{0}}\;\mathrm{Tr}\Big(V_{\mathrm{dis}} (\hat{k},\hat{k}') F(\vec{k}',ik_{n})V_{\mathrm{dis}}^{*} (\hat{k},\hat{k}')\hat{\Delta}_{0}^{\dagger}\Big),
\end{eqnarray}
at the critical temperature ($\Delta_{0}\rightarrow 0$).
Here ($G,F$) are the diagonal and off-diagonal blocks of the Green's function, $\mathcal{G}_{+}$, and one can show that 
\begin{eqnarray}
\lim_{\Delta_{0}\rightarrow 0} G(\vec{k},ik_{n})= -\frac{i\omega_n +\xi_{\vec{k}}\;\sigma_{z}}{\omega_n^2+\xi_{\vec{k}}^{2}}, \quad \lim_{\Delta_{0}\rightarrow 0}\frac{F(\vec{k},ik_{n})}{\Delta_{0}}= -\frac{\psi_{s}(\vec{k})i\sigma_y}{\omega_n^2+\xi_{\vec{k}}^{2}}.
\end{eqnarray}
Adopting the Born approximation, we find that
\begin{eqnarray}
\Sigma_{\omega}=-\frac{i\omega_n}{|\omega_n|}\frac{\Gamma_{l}}{(2l+1)^2}\; y_{m}^{l}(\hat{k})y_{m'}^{l*}(\hat{k})\; \big\langle 1\big \rangle_{m'm},\quad
\frac{\Sigma_{\Delta}}{\Delta_{0}}=\frac{1}{|\omega_n|}
\frac{\Gamma_{l}}{(2l+1)^2}\;y_{m}^{l}(\hat{k})y_{m'}^{l*}(\hat{k})\big\langle \psi_{s}\big \rangle_{m'm}\big[1-\frac{F_{C,0}}{4}\big],\nonumber
\end{eqnarray}
thus, the renormalized parameters are
\begin{eqnarray}
\tilde{\omega}_n&=&\omega_n +\frac{\tilde{\omega}_n}{|\tilde{\omega}_n|}\frac{\Gamma_{l}}{(2l+1)^2}\; y_{m}^{l}(\hat{k})y_{m'}^{l*}(\hat{k})\; \big\langle 1\big \rangle_{m'm} , \label{se52}
\end{eqnarray} 
and
\begin{eqnarray}
\tilde{\Delta}_{0}=\Delta_{0}+\frac{\tilde{\Delta}_{0}}{|\tilde{\omega}_n|}\frac{\Gamma_{l}}{(2l+1)^2}\;y_{m}^{l}(\hat{k})y_{m'}^{l*}(\hat{k})\;\frac{\big\langle \psi_{s}\big \rangle_{m'm}}{\psi_{s}(\hat{k})\ }\big[1-\frac{ F_{C,0}}{4}\big],\label{se53}
\end{eqnarray}
where the weighted average on the Fermi-surface, $\langle \mathcal{O} \rangle_{mm'}\equiv \int_{FS}\frac{ d\Omega}{4\pi} y^{l}_{m}y^{l*}_{m'}\mathcal{O}$, and normalized spherical Harmonics, $y^{l}_m=\sqrt{4\pi}Y^{l}_{m}$ are used with the angular momentum quantum numbers $(l,m)$.
The scattering rate of spin-independent and dependent channels $(\Gamma_{l,v},\Gamma_{l,u})$ are expressed in terms of the DOS of the normal state, $N(0)$,
\begin{eqnarray}
\Gamma_{l,v}=\pi N(0)n_{\mathrm{imp}}v_{l}^{2},\quad\Gamma_{l,u}= \pi N(0)S(S+1)n_{\mathrm{imp}}u_{l}^{2},
\end{eqnarray}
where $S\equiv \sqrt{\vec{S}\cdot \vec{S}}$ is the magnitude of spin of magnetic impurities.
The superconducting fitness function ($F_{C,0}$) is defined with a modified commutator of the pairing and disorder matrices ($\hat{\Delta}_{a},\hat{V}_{\mathrm{dis}}$),
\begin{eqnarray}
F_{C,a}=\mathrm{Tr}\left(\left|\hat{V}_{\mathrm{dis}} \hat{\Delta}_{a}-\hat{\Delta}_{a}\hat{V}_{\mathrm{dis}}^{*}\right|^2\right).
\end{eqnarray}
Here we use $ \hat{\Delta}_{a}=i\sigma_{a}\sigma_y$ and $\hat{V}_{\mathrm{dis}}= I_{2\times 2} \;(\hat{V}_{\mathrm{dis}}=\frac{\vec{S}\cdot \vec{\sigma}}{S})$ for spin-independent (dependent) channels.

\subsection{Linearized gap equation with disorder}\label{ss4_3}
We further evaluate the  critical temperature in the presence (absence) of disorder, $T_{c}\;(T_{c,0})$.
For example, the gap equation of singlet pairing as modified by impurity scattering becomes
\begin{eqnarray}
\frac{1}{2g}=\frac{T_{c,0}}{\mathcal{V}}\sum_{\vec{k},\omega_n}\frac{|\psi_{s}|^{2}}{\omega_n^{2}+\xi_{\vec{k}}^{2}}\rightarrow  \frac{\Delta_0}{2g}=\frac{T_{c}}{\mathcal{V}}\sum_{\vec{k},\omega_n}\frac{|\psi_{s}|^{2}}{\tilde{\omega}_n^{2}+\xi_{\vec{k}}^{2}}\tilde{\Delta}_{0} ,
\end{eqnarray}
where $g>0$ is an attractive pairing interaction.
By replacing the sum over momentum by an integral near the Fermi-energy, we can rewrite the expression
\begin{eqnarray}
\frac{1}{2g}\simeq \pi N(0)T_{c,0}\!\!\sum_{|\omega_n|<\Lambda_{UV}}\!\!\frac{\big\langle|\psi_{s}|^{2}\big\rangle_{\Omega}}{|\omega_n|}\rightarrow  \frac{\Delta_0}{2g}\simeq \pi N(0)T_{c}\!\!\sum_{|\omega_n|<\Lambda_{UV}}\!\!\Big\langle|\psi_{s}|^{2}\frac{\tilde{\Delta}_{0}}{|\tilde{\omega}_n|}\Big\rangle_{\Omega}, \nonumber
\end{eqnarray}
where the UV-energy cutoff ($\Lambda_{UV}$) is imposed on the Matsubara frequencies. 
The angle average on the Fermi-surface, $\langle\mathcal{O} \rangle_{\Omega}=\int_{FS}\frac{d\Omega}{4\pi}\mathcal{O}$, is introduced. 
Substituting the renormalized parameters, $(\tilde{\omega}_n,\tilde{\Delta}_{0})$, we can show that
\begin{eqnarray}
\frac{1}{2gN(0)}&\simeq &2\pi T_{c}\sum_{\omega_n=0}^{\Lambda_{UV}}\!\Big[\frac{\langle|\psi_{s}|^2\rangle_{\Omega}}{\omega_n}-\frac{ \Gamma_{l}}{\omega_n^2}\frac{\alpha_l}{(2l+1)^2}\Big], \label{se56}
\end{eqnarray}
to lowest order in $\Gamma_{l}$ with 
\begin{eqnarray}
\alpha_{l}=\sum_{m,m'}\Big[\langle|\psi_{s}|^{2}\rangle_{mm'} \delta_{m'm}\!-\big\langle \psi_{s}^{*}\big \rangle_{mm'}\big\langle \psi_{s}\big \rangle_{m'm}
\big[1- \frac{F_{C,0}}{4}\big]\Big]/
\langle|\psi_{s}|^2\rangle_{\Omega},\label{se57}
\end{eqnarray}
for $\Gamma_{l}\ll T_{c,0}\ll \Lambda_{UV} $. 
To complete the Matsubara frequency summation in Eq.(\ref{se56}), we can use the useful identities of the digamma function, $\Psi(z)$, and its derivative, $\Psi^{(1)}(z)\equiv d\Psi(z)/dz$,
\begin{eqnarray}
\Psi(z)=-\xi-\sum_{n=0}^{\infty}\Big[ \frac{1}{n+z}-\frac{1}{n+1}\Big],\quad \Psi^{(1)}(z)=\sum_{n=0}^{\infty}\; \frac{1}{(n+z)^2},\nonumber
\end{eqnarray}
where $\xi$ is the Euler constant. 
The approximations, $\Psi(z)\simeq \log z$, $\Psi^{(1)}(z)\simeq \frac{1}{z}$, for a large $|z|$ yield that 
\begin{eqnarray}
2\pi T_{c}\sum_{\omega_n=0}^{\Lambda_{UV}}\frac{1}{\omega_n}=-\Big[ \Psi\left( \frac{1}{2}\right) -\Psi\left( \frac{1}{2}+\frac{\Lambda_{UV}}{2\pi T_{c}}\right) \Big]\simeq -\Big[  \Psi\left( \frac{1}{2}\right) -\log\Big( \frac{\Lambda_{UV}}{2\pi T_c}\Big)\Big],\nonumber
\end{eqnarray}
and
\begin{eqnarray}
2\pi T_{c}\sum_{\omega_n=0}^{\Lambda_{UV}}\frac{1}{\omega_n^2}=\frac{1}{2\pi T_c}\Big[ \Psi^{(1)}\left( \frac{1}{2}\right) -\Psi^{(1)}\left( \frac{1}{2}+\frac{\Lambda_{UV}}{2\pi T_{c}}\right) \Big]\simeq \frac{1}{2\pi T_c} \Big[ \frac{\pi^2}{2}-\frac{2\pi T_{c}}{\Lambda_{UV}}\Big],\nonumber
\end{eqnarray}
for $ T_{c}\ll \Lambda_{UV} $.
Thus, the critical temperature is  
\begin{eqnarray}
\log\left(\frac{T_{c}}{T_{c,0}}\right)=-\frac{\pi}{4(2l+1)^2}\frac{\alpha_{l,v(u)}
}{T_{c,0}}\Gamma_{l,v(u)}, \label{setc1}
\end{eqnarray}
where the parameter $\alpha_{l}$ quantifies the fragility of a superconducting state under disorder.

Assuming SO(3) symmetry, the singlet gap function can be chosen as, $\psi_{s}=y^{L}_{0}$, with an even number, $L$.
We further simplify the expression for $\alpha_{l}$ by using a selection rule, $\langle y^{L}_{M}\rangle_{m,m'}\propto \delta_{m-m'+M}$, and find its lower bound
\begin{eqnarray}
\alpha_{l}=\sum_{m}\Big[\langle|\psi_{s}-\langle \psi_s\rangle_{mm}|^{2}\rangle_{mm} +|\big\langle \psi_{s}\big \rangle_{mm}\big|^{2}  \frac{F_{C,0}}{4} \Big]/\langle|\psi_{s}|^2\rangle_{\Omega}\geq 0.
\end{eqnarray}
Here, the equality holds if the two quantities ($\langle|\psi_{s}-\langle \psi_s\rangle_{mm}|^{2}\rangle_{mm},F_{C,0}$) become zero at the same time.
The first term, $\langle| \psi_{s}-\langle\psi_s\rangle_{mm}|^2\rangle$, is a variance of a gap function, thus evidently can be zero only for an $s$-wave pairing.
The second term, $F_{C,0}$, always vanishes for spin-independent disorder.

\begin{table}[t]
\centering 
\renewcommand{\arraystretch}{1.3}
\begin{tabular}{C{0.08\linewidth}C{0.15\linewidth}C{0.12\linewidth}C{0.15\linewidth}C{0.12\linewidth}}\hline \hline
$l$&  \thead{(A) BG-FS\\ ($\psi_{s}=d_{xz}+id_{yz}$)}&  \thead{(B) Line-node\\ ($\psi_{s}=d_{xz}$)}& \thead{(C) Point-node\\ ($\vec{\psi}_{t}=(p_x+ip_y)\hat{z}$)}&\thead{(D) Full gap\\ ($\psi_{s}=s$)}
\\ \hline
$0$&$(1,1)$& $(1,1)$&$(1,1)$& $(0,2)$\\ 
$1$&$(\frac{9}{5},\frac{21}{5})$&$(\frac{9}{5},\frac{21}{5})$&$(3,3)$& $(0,6)$ \\ 
$2$&$(\frac{25}{7},\frac{45}{7})$& $(\frac{25}{7},\frac{45}{7})$&$(5,5)$& $(0,10)$\\ 
$3$&$(\frac{77}{15},\frac{133}{15})$&$(\frac{77}{15},\frac{133}{15})$&$(7,7)$& $(0,14)$ \\
$4$&$(\frac{513}{77},\frac{873}{77})$&$(\frac{513}{77},\frac{873}{77})$& $(9,9)$& $(0,18)$\\ 
\hline \hline
\end{tabular}
\caption{
List of the parameters $(\alpha_{l,v},\alpha_{l,u})$ of four different nodal superconducting states. 
The five lowest order of impurity potential channels, $l=0,\cdots, 4$ is tabulated. 
We consider the four specific gap functions, $\psi_{s}=d_{xz}+id_{yz}$,$\psi_{s}=d_{xz}$,$\vec{\psi}_{t}=(p_{x}+ip_{y})\hat{z}$, $\psi_{s}=s$, which realize a BG-FS, line-node, point-node, full-gap, respectively. 
}\label{st2}
\end{table}
\begin{figure}[t]
\includegraphics[scale=.5]{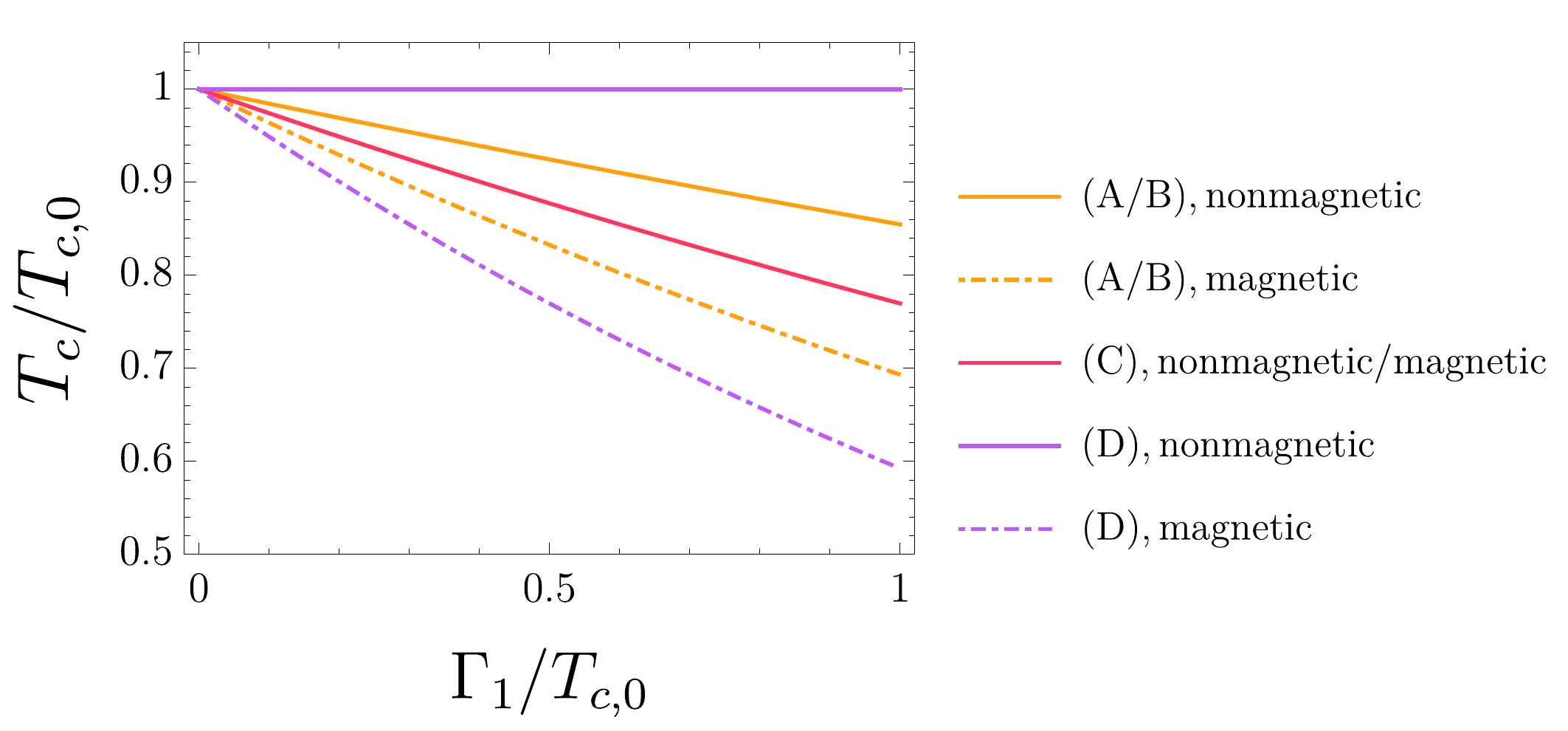}\vspace{-10pt}
\caption{
The scattering rate dependence of critical temperature $(T_{c})$ of superconductors with four different nodes (Eq.(\ref{setc1}, \ref{setc2})).
We illustrate each case by choosing a specific pairing state (See Table.\ref{st2}) and a $l=1$ impurity potential. 
Orange, pink, purple colors denote a BG-FS(A)/ line-node(B), point-node (B), full-gap (D)
and solid and dotted lines denote non-magnetic and magnetic impurities, respectively. 
While an s-wave superconductor is perfectly robust under non-magnetic impurities, other superconductors including a BG-FS are affected by disorder. 
The reason why the results of line-node and BG-FS are the same is related to the SO(3) symmetry, which is an artifact of the fine-tuned case. 
We emphasize that the critical behavior is not universal behavior but depends on microscopic details. 
}\label{sf7}
\end{figure}

\subsection{Generalizations}\label{ss4_5}
One can generalize Eq. (\ref{se57},\ref{setc1}) into the case with lower symmetry or triplet pairing. 
The former case may be achieved by replacing the quantum numbers $(l,m)$ with an irreducible representation index (R) of a given point group.
To reach the latter case, we consider a pseudospin triplet pairing, 
\begin{eqnarray}
\Delta_{+}(\vec{k})=\Delta_{0}(\vec{\psi}_{t}(\vec{k})\cdot\vec{\sigma})i
\sigma_y,
\end{eqnarray}
where $\vec{\psi}_{t}(\vec{k})$ is a d-vector for spin triplet superconductors \cite{s_mineev}.  
The parameter $\alpha_{l}$ for triplet pairing is
\begin{eqnarray}
\alpha_{l}=\sum_{a,b=1}^{3}\sum_{m,m'}\Big[\langle \psi_{t,a}^{*}\psi_{t,b}\rangle_{mm'} \delta_{m'm}\delta_{ab}-\big\langle \psi_{t,a}^{*}\big \rangle_{mm'}\big\langle \psi_{t,b}\big \rangle_{m'm}
\big[\delta_{ab}- \frac{F_{C,ab}}{4}\big]\Big]/\langle|\vec{\psi}_t|^2\rangle_{\Omega}, 
\end{eqnarray}
with
\begin{eqnarray}
F_{C,ab}=\mathrm{Tr}\left( (\hat{V}_{\mathrm{dis}} \hat{\Delta}_{a}-\hat{\Delta}_{a}\hat{V}_{\mathrm{dis}}^{*})^{\dagger} (\hat{V}_{\mathrm{dis}} \hat{\Delta}_{b}-\hat{\Delta}_{b}\hat{V}_{\mathrm{dis}}^{*})\right), 
\end{eqnarray}
and
\begin{eqnarray}
\log\left(\frac{T_{c}}{T_{c,0}}\right)=-\frac{\pi}{4(2l+1)^2}\frac{\alpha_{l,v(u)}
}{T_{c,0}}\Gamma_{l,v(u)}. \label{setc2}
\end{eqnarray}

\subsection{Comparison of BG-FS and other superconductors }\label{ss4_4}
We here consider a SO(3) symmetric normal Fermi-surface with two different gap functions: $i)$ $\psi_{s}=y^{0}_{0}\propto 1$, $ii)$ $\psi_{s}= y^{2}_{1}\propto(k_{x}+ik_y) k_{z}$.
The former case is a conventional $s$ -wave pairing 
and the latter case is a TRSB chiral pairing, which corresponds to a BG-FS, as in the previous section.

The parameter $\alpha_{l}$ of an $s$-wave pairing is proportional to the superconducting fitness function.
We evaluate the exact values, $(\alpha_{l,v}=0,\alpha_{l,u}=4l+2)$, for both spin-independent and dependent channels, where the original Anderson theorem is manifested. 
For a chiral pairing, the superconducting fitness function is not enough to measure the robustness of the superconducting state.
In Table. \ref{st2} and Fig. \ref{sf7},  we tabulate the values of $(\alpha_{l,v},\alpha_{l,u})$ of the five lowest order channels and plot their behivors especially for $l=1$ case even considering line-nodal and point nodal superconductors. 
It is shown that a BG-FS is fragile even under non-magnetic disorder, in contrast to s-wave pairing.

\bibliographystyle{apsrev4-1}
%

\end{document}